\documentclass{elsart}
\usepackage{amsmath,latexsym,psfrag,caption2}
\usepackage{graphicx}
\usepackage{cite}
\usepackage{srcltx}
\usepackage{axodraw}
\newcounter{bla}

\newcounter{im}
\newcommand{\example}{\stepcounter{im}\includegraphics[scale=0.9]{HypExp2Examples_\arabic{im}.eps}}
\def\exboxlength{1.0\textwidth}

\begin{document}
\begin{frontmatter}
\hfill PITHA-07/06\\ \hfill SLAC-PUB-12748\\
\title{{\tt HypExp 2}, \\ Expanding Hypergeometric Functions about Half-Integer Parameters}

\author[a]{Tobias Huber},
\author[b,c]{Daniel Ma\^{\i}tre},


\address[a]{Institut f\"{u}r Theoretische Physik E, RWTH Aachen,
D-52056 Aachen, Germany}
\address[b]{Stanford Linear Accelerator Center,2575 Sand Hill,Menlo Park, CA 94025}
\address[c]{Institut f\"ur Theoretische Physik,University of Z\"urich, Winterthurerstrasse 190, CH-8057 Z\"urich}

\begin{abstract}
In this article, we describe a new algorithm for the expansion of hypergeometric functions about half-integer parameters. The
implementation of this algorithm for certain classes of hypergeometric functions in the already existing {\tt Mathematica}  package {\tt
HypExp} is described. Examples of applications in Feynman diagrams with up to four loops are given.

\begin{flushleft}
PACS:02.10.De, 02.30.Gp 

\end{flushleft}

\begin{keyword}
Hypergeometric function; expansion; half integers; Mathematica.
\end{keyword}

\end{abstract}

\end{frontmatter}


{\bf NEW VERSION PROGRAM SUMMARY}

\begin{small}
\noindent
{\em Manuscript Title: }    {\tt HypExp 2}, Expanding Hypergeometric Functions about Half-Integer Parameters          \\
{\em Authors:} Tobias Huber, Daniel Ma\^{\i}tre                                                \\
{\em Program Title:} HypExp                                         \\
{\em Journal Reference:}                                      \\
{\em Catalogue identifier:}                                   \\
{\em Licensing provisions:} None                                  \\
{\em Programming language:} Mathematica                                  \\
{\em Computer:} Computers running Mathematica                                              \\
{\em Operating system:} Linux, Windows, Mac                                      \\
{\em RAM:} Depending on the complexity of the problem                                              \\
{\em Number of processors used:}                              \\
{\em Supplementary material:}                                 \\
{\em Keywords:} Hypergeometric function, expansion, half integers, Mathematica  \\
{\em PACS:}  02.10.De, 02.30.Gp                               \\
{\em Classification:}  CPC 4.7                                       \\
{\em External routines/libraries:}                                      \\
{\em Subprograms used:} The package uses the package HPL included in the distribution                                      \\
{\em Catalogue identifier of previous version:} ADXF\_v1\_0              \\
{\em Journal reference of previous version:} Comput. Phys. Commun. 175(2006)122                 \\
{\em Does the new version supersede the previous version?:} Yes   \\

{\em Nature of the problem:}\\
  Expansion of hypergeometric functions about parameters that are integer and/or half-integer valued.
   \\
{\em Solution method:}\\
New algorithm implemented in Mathematica.  
   \\
{\em Reasons for the new version:}\\
  Expansion about half-integer parameters.
   \\
{\em Summary of revisions:}\\ Ability to expand about half-integer valued parameters added.   
   \\
{\em Restrictions:}\\
The classes of hypergeometric functions with half-integer parameters that can be expanded are listed in the long write-up.  
   \\
{\em Unusual features:}\\
   \\
{\em Additional comments:}\\
   \\
{\em Running time:}\\
Depending on the expansion.  
   \\
{\em References:}

\end{small}

\newpage


\hspace{1pc}
{\bf LONG WRITE-UP}

\renewcommand{\d}{\mathrm{d}}
\newcommand{\arctanh}{\mathrm{arctanh}}
\newcommand{\arccot}{\mathrm{arccot}}
\newcommand{\arccsc}{\mathrm{arccsc}}
\newcommand{\sgn}{\mathrm{sgn}}
\newcommand{\prim}[1]{{#1^{\prime}}}
\newcommand{\tr}{\mathrm{Tr}}
\newcommand{\naeher}{\!\!\!}
\newcommand{\vect}[1]{\vec #1}
\newcommand{\ddd}[1]{\d\vect{#1} }
\def\be{\begin{equation}}
\def\ee{\end{equation}}
\def\bea{\begin{eqnarray}}
\def\eea{\end{eqnarray}}
\def\nnb{\nonumber}
\def\eps{\epsilon}
\def\dps{\displaystyle}
\newcommand{\hs}[1]{\hspace*{#1 pt}}
\newcommand{\vs}[1]{\vspace*{#1 pt}}
\newcommand{\MB}[2]{\hs{-12} \int\limits_{\hs{15}_{ #1 \hs{-10}-i \,
\infty}}^{\hs{15}^{ #1 +i\, \infty}} \hs{-15} \frac{d #2}{2\pi i}}
\newcommand{\pFq}[5]{\, \! _{#1} F_{#2}( #3 \, ; \, #4 \, ; \, #5 )}

\maketitle
\section{Introduction}
Hypergeometric functions ${}_PF_Q$ appear in many branches of science. They appear, in particular, in particle physics during the
calculation of radiative corrections to scattering cross sections in loop~\cite{loop1,loop2,loop3,Anastasiou:1999bn,Fleischer:2003rm,loop5,loop6,Gehrmann:2006wg} or
phase space\cite{phasespace1,phasespace2,Huber:2005ig} integrals. In the context of dimensional regularization, the parameters of the
hypergeometric functions are usually functions of an arbitrary space-time dimension $D=4-2 \epsilon$ where $\epsilon$ regulates infrared
and/or ultraviolet divergences. For physical observables, only the limit $\epsilon\rightarrow 0$ is of importance. Because divergences
appear in form of poles $1/\epsilon^n$, not only the constant term of the Taylor expansion around the physical dimension is of physical
relevance. For this reason one is often confronted with the task of expanding hypergeometric functions about their parameters. 

Systematic approaches to the expansion of hypergeometric functions have been developed
\cite{weinzierl1,Weinzierl:2004bn,Kalmykov:2007pf} and have been implemented in GiNaC~\cite{weinzierl2,GINAC},
Mathematica~\cite{Huber:2005yg}, and FORM~\cite{Moch:2005uc}. These implementations,
however, are restricted to the expansion of hypergeometric functions about integer-valued parameters.

In computations involving massive particles~%
\cite{Davydychev:1992mt,Broadhurst:1993mw,Tarasov:2006nk,Davydychev:1998si,Fleischer:2003rm,Jegerlehner:2002em,Jegerlehner:2003py,Davydychev:2000na,Davydychev:2003mv,Schroder:2005va,Bejdakic:2006vg,Grozin:2007ap,Argeri:2007up} the
hypergeometric functions can contain half-integer parameters. Methods have been developed to expand hypergeometric functions with
half-integer parameters~\cite{Weinzierl:2004bn,Kalmykov:2006pu,Kalmykov:2006hu,Kalmykov:2007dk}. However, none of these methods has been made available
in a user-friendly package written in a widely used computer algebra programme.

The aim of this work is to introduce a new method for expanding hypergeometric functions about half-integer parameters, and its
implementation in the for this purpose extended version of the publicly available Mathematica~\cite{Mathematica} package {\tt
HypExp}~\cite{Huber:2005yg}.

This paper is articulated as follows. In section~\ref{sec:halfhypexp}, which is quite technical, we introduce the algorithm that allows
the expansion of hypergeometric functions about half-integer parameters. Section~\ref{manual} describes the implementation of this
algorithm in the existing Mathematica package {\tt HypExp}. Section~\ref{sec:examples} contains examples of Feynman
diagrams to which the package can be applied, and we conclude in section~\ref{sec:conclusions}. Appendix~\ref{sec:analytic} is devoted
to the analytic continuation of the expansion, and appendix~\ref{sec:basis} contains the computation of the basis functions for several
types of hypergeometric functions.
\section{Expansion of hypergeometric functions about half-integer parameters}\label{sec:halfhypexp}
In this section we describe an algorithm to expand hypergeometric functions about half-integer parameters. The strategy of the algorithm
will be to express the expansion of a hypergeometric function of a given type in terms of certain operators acting on the expansion of one hypergeometric function of the
same type, the latter will be referred to as the basis function of this type.  We will then deal with the task of expanding the basis function and carrying out the aforementioned
operations. Our approach differs from other reduction algorithms \cite{Kalmykov:2007pf,takayama,Kalmykov:2006hu,Kalmykov:2007dk,Gauss,Yoshida,Iwasaki,Koornwinder,Vidunas,Kalmykov:2006pu} in that we relate only expansions to a
given order with the expansion of the basis function, and not the functions themselves.

We will start this section by giving some definitions and notation, before we explain the reduction to the basis function by means of an
example, namely by considering a certain type of a ${}_2F_1$ function. We will then extend this method and show how to reduce a general
function ${}_PF_{P-1}$ to the respective basis function. Afterwards, we explain a strategy to find the all-order expansion of the basis
functions and give some applications.

\subsection{Definitions and notation}\label{sec:definitions}
In this subsection we introduce some definitions and notation that will be used in the following sections. In this paper we consider
hypergeometric functions (hereafter HF)
\[{}_{P}F_{P-1}(A_1,...,A_{P};B_1,...,B_{P-1};z),\] 
whose parameters $A_j$ and $B_j$ have the following form 
\[i+\gamma\epsilon \quad\mbox{ or }\quad i+\frac{1}{2}+\gamma\epsilon\;,\]
where $i$ is an integer, $\gamma$ is a real coefficient, and $\epsilon$ is the parameter in which we will expand the HF.

We will denote the finite part (i.e.\ either $i$ or $i+\frac{1}{2}$) of a parameter by lower case Latin letters, the coefficients of the
expansion parameter $\epsilon$ will be labeled by Greek $\alpha$ or $\beta$, and the parameter as a whole will be referred to by capital
$A$ or $B$. $A$, $a$, and $\alpha$ always correspond to an element of the first subset of parameters whereas $B$, $b$, and $\beta$
correspond to the second subset of parameters.  

We will refer to an HF as being of the type $P^i_j$ if $i$ out of the $P$ $a_k$'s and $j$ out of the $P-1$ $b_k$'s are half-integers,
and the other $a_k$'s and $b_k$'s are integers. 

We define the short-hand notation
\be\label{eq:a:bproducts}
\prod\limits_{j}^{a:a}=1,\quad\prod\limits_{j}^{a:b}f(j)=\prod\limits_{j=a}^{b-1}f(j) \quad\textnormal{if}\quad a<b,\quad\prod\limits_{j}^{a:b}f(j)=\prod\limits_{j=b}^{a-1}\displaystyle\frac{1}{f(j)} \quad\textnormal{if}\quad a>b\;,
\ee
so that
\be
\Gamma(b)=\Gamma(a)\prod\limits_{j}^{a:b}(j)\;.
\ee
\subsection{Algorithm}
The algorithm is based on the fact that one can write fractions of the type
\[\frac{x^i}{(i+j)^n}\]
with $i+j\not = 0$ in terms of the integration operator
\begin{eqnarray}
J^+(j)[f](x)&=&\frac{1}{x^j}\int\limits_0^x\d x' x'^{j-1}f(x')\equiv J^+(j,1)[f](x),\nnb\\
J^+(j,n)[f](x)&\equiv&\left(J^+(j)\right)\left[J^+(j,n-1)[f]\right](x)
\end{eqnarray} 
in the following way
\begin{eqnarray}
\frac{x^i}{i+j}&=&\frac{1}{x^j}\int\limits_0^x\d x' x'^{j-1} x'^i\equiv J^+(j)[y^i](x),\\
\frac{x^i}{(i+j)^n}&
\equiv& J^+(j,n)[y^i](x)\;.
\end{eqnarray}
Polynomials in $i$ and $x$ can also be re-written using 
\[i x^i=x\frac{\d}{\d x} x^i,\qquad i^n x^i=\left(x\frac{\d}{\d x}\right)^n x^i\;,\]
for which we define the operators
\begin{eqnarray}
J^-(j)[f](x)&=&\frac{1}{x^{j-1}}\frac{\d}{\d x}x^jf(x)\equiv J^-(j,1)[f](x)\;,\nnb\\
J^-(j,n)[f](x)&\equiv&\left(J^-(j)\right)\left[J^-(j,n-1)[f]\right](x)\; .
\end{eqnarray}
We then have
\be
i x^i = J^-(0)[y^i](x) \; , \qquad i^n x^i = J^-(0,n)[y^i](x) \; .
\ee
We will use these facts to express the expansion of an HF in terms of these differentiation and integration operators acting on another
HF of the same type, the latter will be referred to as the basis function of this type. In the following section, we will illustrate the
algorithm with an HF of type $2^1_1$.
\subsubsection{Warm up: Hypergeometric functions of type $2^1_1$}
We start from the definition of the hypergeometric function ${}_2F_1$ 
\be\label{eq:serdef}
{}_2F_{1}(A_1, A_2;B_1;x)=1+\frac{\Gamma(B_1)}{\Gamma(A_1)\Gamma(A_{2})}\;\sum\limits_{i=1}^\infty\frac{\Gamma(A_1+i)\Gamma(A_2+i)}{\Gamma(B_1+i)\Gamma(i+1)}x^i\;.
\ee
We consider the case where the finite parts of $A_1$ and $B_1$ are half-integers and that of $A_2$ is an integer
\[A_1=a_1+\frac{1}{2}+\alpha_1\epsilon,\quad A_2=a_2+\alpha_2 \epsilon,\quad B_1=b_1+\frac{1}{2}+\beta_1\epsilon\;.\]
Using the relation $x\Gamma(x)=\Gamma(x+1)$ we can transform the expression (\ref{eq:serdef}) into
\begin{eqnarray}\label{serdef2}
\lefteqn{{}_2F_{1}(a_1+\frac{1}{2}+\alpha_1\epsilon, a_2+\alpha_2\epsilon;b_1+\frac{1}{2}+\beta_1\epsilon;x)=}&&\nonumber\\
&=&1+
\frac{\Gamma(\frac{1}{2}+\beta_1\epsilon)}{\Gamma(\frac{1}{2}+\alpha_{1}\epsilon)\Gamma(1+\alpha_2\epsilon)}\frac{\prod\limits_{j}^{0:b_1}(j+\frac{1}{2}+\beta_1\epsilon)}{\prod\limits_{j}^{0:a_1}(j+\frac{1}{2}+\alpha_1\epsilon)\prod\limits_{j}^{1:a_2}(j+\alpha_2\epsilon)}\nonumber\\
&\times&\sum\limits_{i=1}^\infty\underbrace{\frac{\prod\limits_{j}^{0:a_1}(i+j+\frac{1}{2}+\alpha_1\epsilon)\prod\limits_{j}^{1:a_2}(i+j+\alpha_2\epsilon)}{\prod\limits_{j}^{0:b_1}(i+j+\frac{1}{2}+\beta_1\epsilon)}}_{D}\frac{\Gamma(i+\frac{1}{2}+\alpha_1\epsilon)\Gamma(i+1+\alpha_2\epsilon)}{\Gamma(i+\frac{1}{2}+\beta_1\epsilon)\Gamma(i+1)}x^i\;,\nonumber\\
\end{eqnarray}
where we made use of the short-hand notation defined in Eq.~(\ref{eq:a:bproducts}). The factor $D$ in Eq.~(\ref{serdef2}) can be turned into
partial fractions with respect to $i$. This yields a sum of factors $i^n$,  $1/(i+j+\gamma\epsilon)^n$ or
$1/(i+j+1/2+\gamma\epsilon)^n$ .
\begin{eqnarray}\label{eq:2F1partialfrac}
\lefteqn{{}_2F_{1}(a_1+\frac{1}{2}+\alpha_1\epsilon, a_2+\alpha_2\epsilon;b_1+\frac{1}{2}+\beta_1\epsilon;x)=}&&\nonumber\\&=&
1+\frac{\prod\limits_{j}^{0:b_1}(j+\frac{1}{2}+\beta_1\epsilon)}{\prod\limits_{j}^{0:a_1}(j+\frac{1}{2}+\alpha_1\epsilon)\prod\limits_{j}^{1:a_2}(j+\alpha_2\epsilon)}\frac{\Gamma(\frac{1}{2}+\beta_1\epsilon)}{\Gamma(\frac{1}{2}+\alpha_1\epsilon)\Gamma(1+\alpha_{2}\epsilon)}\nonumber\\
&&\times\sum\limits_{i=1}^\infty\left(\sum\limits_{j\ge0,n} \frac{C_{j,n}^+}{(i+j+\gamma\epsilon)^{n}}+\sum\limits_{j<0,n} \frac{C_{j,n}^+}{(i+j+\gamma\epsilon)^{n}}\right.\nonumber\\
&&\qquad\qquad\qquad\qquad\qquad\qquad\qquad+\left.\sum\limits_{j,n} \frac{C_{j,n}^{1/2}}{(i+\frac{1}{2}+j+\gamma\epsilon)^{n}}+\sum_n C_n^-i^n\right)\nonumber\\
&&\times\frac{\Gamma(i+1+\alpha_1\epsilon)\Gamma(i+\frac{1}{2}+\alpha_2\epsilon)}{\Gamma(i+\frac{1}{2}+\beta_1\epsilon)\Gamma(i+1)}x^i\;,
\end{eqnarray}
where the coefficients $C_{j,n}^{\pm}$ and $C^{1/2}_{j,n}$ are polynomials in $\epsilon$.
We will now show in turn for each of the four double sums how it can be expressed in terms of integration and differentiation operators
acting on a basis function $B$.

In the sum over positive $j$'s
the denominators can be expanded in $\epsilon$, which yields factors of $1/(i+j)^n$ that can be expressed as integration operators
\begin{equation}
J^+(j,n)=\left(\frac{1}{x^j}\int\limits_0^x\d x' x'^{j-1}\right)^n 
\end{equation}
acting on 
\begin{eqnarray}\label{eq:b11}
B&\equiv&\frac{\Gamma(\frac{1}{2}+\beta_1\epsilon)}{\Gamma(\frac{1}{2}+\alpha_1\epsilon)\Gamma(1+\alpha_2\epsilon)}\sum\limits_{i=1}^\infty\frac{\Gamma(i+\frac{1}{2}+\alpha_1\epsilon)\Gamma(i+1+\alpha_2\epsilon)}{\Gamma(i+\frac{1}{2}+\beta_1\epsilon)\Gamma(i+1)}x^i\nonumber\\
&=&{}_2F_1(\frac{1}{2}+\alpha_1\epsilon,1+\alpha_2\epsilon,\frac{1}{2}+\beta_1,x)-1\;.
\end{eqnarray}
The same applies for the third sum since it contains terms of the form \\$1/(i+\frac{1}{2}+j+\gamma\epsilon)^n$.
They can be expressed using the integration operator
\[J^+_{1/2}(j,n)=\left(\frac{1}{x^{j+\frac{1}{2}}}\int\limits_0^x\d x' x'^{j-\frac{1}{2}}\right)^n=J^+(j+\frac{1}{2},n)\;.\] 
The factors $i^n$ in the last sum can be written as differentiation operators
\begin{equation}
J^-(0,n)=\left(x\frac{\d}{\d x}\right)^n
\end{equation}
acting on $B$. The second of the above sums requires more work. For these terms which have $j=-k<0$ we decompose the outer sum over $i$
in two parts, one running from 1 to $k$, and the other one running from $k+1$ to infinity, according to the following
structure,
\be
\sum\limits_{i=1}^\infty \frac{x^i}{(i-k+\gamma\epsilon)^n}B_i=\sum\limits_{i=1}^{k} \frac{x^i}{(i-k+\gamma\epsilon)^n}B_i+\sum\limits_{i=k+1}^\infty\frac{x^i}{(i-k+\gamma\epsilon)^n}B_i\;.
\ee
The last term can be worked out as follows
\begin{eqnarray}
\lefteqn{\sum\limits_{i=k+1}^\infty\frac{x^i}{(i-k+\gamma\epsilon)^n}B_i}&&\nonumber\\
&=&\sum\limits_{i=1}^\infty\frac{x^i
x^k}{(i+\gamma\epsilon)^n}B_{i+k}=x^k\sum\limits_{i=1}^\infty\frac{x^i}{(i+\gamma\epsilon)^n}B_{i+k}\nnb\\
&=&x^k\sum\limits_{l=0}^\infty\frac{(n+l-1)!}{l!(n-1)!}(-\gamma\epsilon)^l\sum\limits_{i=1}^\infty\frac{x^i}{(i)^{n+l}}B_{i+k}\nonumber\\
&=&x^k\sum\limits_{l=0}^\infty\frac{(n+l-1)!}{l!(n-1)!}(-\gamma\epsilon)^l\left[J^+(0,n+l)\right]\left(\sum\limits_{i=1}^\infty x^{-k} x^i \tilde B_{i,k}\right)\nonumber\\
&=&x^k\sum\limits_{l=0}^\infty\frac{(n+l-1)!}{l!(n-1)!}(-\gamma\epsilon)^l\left[J^+(0,n+l)\right]\left(x^{-k} \tilde B(k,x)\right)\;,
\end{eqnarray}
where
\begin{equation} \tilde B_{i,k}=\left\{\begin{array}{c}0,\qquad i\le k\\B_i,\qquad i>k\end{array}\right.\qquad\qquad\textnormal{and}\qquad\qquad\tilde B(k,x)=\sum\limits_{i=1}^\infty x^i \tilde B_{i,k}\;.
\end{equation}
One can now write the terms with $j<0$ in $D$ as an operator
\begin{eqnarray}
\lefteqn{\sum\limits_{i=1}^{\infty}\frac{x^i}{(i+j+\gamma\epsilon)^n}\, B_i\rightarrow J^+(j,n,\gamma)B}&&\nonumber\\
&\equiv&\sum\limits_{i=1}^{-j}\frac{B_ix^i}{(i+j+\gamma\epsilon)^n}+x^{-j}\sum\limits_{l=0}^\infty
\frac{(n+l-1)!}{l!(n-1)!}(-\gamma\epsilon)^l J^+(0,n+l)\left(x^{j} \tilde B(-j,x)\right)\;,\nonumber\\
\end{eqnarray}
where $B_i$ is the $i$-th coefficient of the Taylor expansion in $x$ of $B$ about $x=0$ 
\begin{equation}
B_i=\frac{\Gamma(a_1+\frac{1}{2}+i+\alpha_1\epsilon)\Gamma(a_2+1+i+\alpha_2\epsilon)\Gamma(b_1+\frac{1}{2}+\beta_1\epsilon)}{\Gamma(a_1+\frac{1}{2}+\alpha_1\epsilon)\Gamma(a_2+1+\alpha_2\epsilon)\Gamma(b_1+\frac{1}{2}+i+\beta_1\epsilon)
i!}\;.
\end{equation}
The final formula reads
\begin{eqnarray}
\lefteqn{{}_2F_{1}(a_1+\frac{1}{2}+\alpha_1\epsilon, a_2+\alpha_2\epsilon,b_1+\frac{1}{2}+\beta_1\epsilon,x)}&&\nnb\\&=&
1+\frac{\prod\limits_{j}^{0:b_1}(j+\frac{1}{2}+\beta_1\epsilon)}{\prod\limits_{j}^{0:a_1}(j+\frac{1}{2}+\alpha_1\epsilon)\prod\limits_{j}^{1:a_2}(j+\alpha_2\epsilon)}\nonumber\\
&\times&\left(\sum\limits_{j\ge0,n} C_{j,n}^+J^+(j,n)+\sum\limits_{j<0,n,\gamma} C_{j,n,\gamma}^+J^+(j,n,\gamma)\right.\nonumber\\
&&\left.\qquad\qquad +\sum\limits_{j,n} C_{j,n}^{1/2}J^+(j+\frac{1}{2},n)+\sum_n C_n^-J^-(0,n)\right)B\;.
\end{eqnarray}
It expresses the HF under consideration in terms of integration and differentiation operators acting on the basis
function $B$ which we have chosen to be as in Eq.~(\ref{eq:b11}). We will now generalize this procedure to arbitrary
${}_PF_{P-1}$.
\subsubsection{General method}
We now consider the general case for the expansion of 
\begin{eqnarray}\label{eq:PFPm1Ser}
\lefteqn{{}_{P}F_{P-1}(A_1,...,A_{P};B_1 ,...,B_{P-1},x)
=}&&\nonumber\\
&&1+\frac{\prod_{j=1}^{P-1}\Gamma(B_j)}{\prod_{l=1}^{P}\Gamma(A_l)}\;\sum\limits_{i=1}^\infty\frac{\prod_{l=1}^{P}\Gamma(A_l+i)}{\prod_{j=1}^{P-1}\Gamma(B_j+i)}\frac{x^i}{\Gamma(i+1)}\;,
\end{eqnarray}
with
\begin{eqnarray}
A_i=a_i+\frac{1}{2}+\alpha_i\epsilon,\quad 1\le i\le r,&\qquad& A_i=a_i+\alpha_i\epsilon,\quad r<i\le P\nonumber\\
B_i=b_i+\frac{1}{2}+\beta_i\epsilon,\quad 1\le i\le s,&\qquad& B_i=b_i+\beta_i\epsilon,\quad s<i\le P-1\;,\nonumber\\
 \end{eqnarray}
where all $a_i$'s and $b_i$'s are integers and the HF is therefore of type $P^r_s$ according to section~\ref{sec:definitions}. Using the
relations
\[x\Gamma(x)=\Gamma(x+1),\qquad \Gamma(x+m)=\Gamma(x+n) \prod\limits_{j}^{n:m}(x+j)\;,\] 
we can transform Eq.~(\ref{eq:PFPm1Ser}) into
\begin{eqnarray}\label{eq:general}
\lefteqn{{}_{P}F_{P-1}(a_1+\frac{1}{2}+\alpha_1 \epsilon,...,a_{P}+\alpha_{P}\epsilon;b_1+\frac{1}{2}+\beta_1
\epsilon,...,b_{P-1}+\beta_{P-1}\epsilon,x)=}&&\nonumber\end{eqnarray}
\begin{eqnarray}
&=&1+
\frac{
\left(\prod\limits_{l=1}^{s}\Gamma(m_l+\frac{1}{2}+\beta_l\epsilon)\right)
\left(\prod\limits_{l=s+1}^{P-1}\Gamma(m_l+\beta_l\epsilon)\right)
}{
\left(\prod\limits_{l=1}^{r}\Gamma(n_l+\frac{1}{2}+\alpha_l\epsilon)\right)
\left(\prod\limits_{l=r+1}^{P}\Gamma(n_l+\alpha_{l}\epsilon)\right)
}\nonumber\\
&\times&\frac{
\left(\prod_{l=1}^{s}\prod\limits_{j_l}^{m_l:b_l}(j_l+\frac{1}{2}+\beta_l\epsilon)\right)
\left(\prod_{l=s+1}^{P-1}\prod\limits_{j_l}^{m_l:b_l}(j_l+\beta_l\epsilon)\right)
}{
\left(\prod\limits_{l=1}^{r}\prod\limits_{j_l}^{n_l:a_l}(j_l+\frac{1}{2}+\alpha_l\epsilon)\right)
\left(\prod\limits_{l=r+1}^{P}\prod\limits_{j_l}^{n_l:a_l}(j_l+\alpha_l\epsilon)\right)
}\nonumber\\
&\times&\sum\limits_{i=1}^\infty\underbrace{
\frac{
\left(\prod\limits_{l=1}^{r}\prod\limits_{j_l}^{n_l:a_l}(i+j_l+\frac{1}{2}+\alpha_l\epsilon)\right)
\left(\prod\limits_{l=r+1}^{P}\prod\limits_{j_l}^{n_l:a_l}(i+j_l+\alpha_l\epsilon)\right)
}{
\left(\prod_{l=1}^{s}\prod\limits_{j_l}^{m_l:b_l}(i+j_l+\frac{1}{2}+\beta_1\epsilon)\right)
\left(\prod_{l=s+1}^{P-1}\prod\limits_{j_l}^{m_l:b_l}(i+j_l+\beta_l\epsilon)\right)
}
}_{D}\nonumber\\
&\times&\frac{
\left(\prod\limits_{l=1}^{r}\Gamma(i+n_l+\frac{1}{2}+\alpha_l\epsilon)\right)
\left(\prod\limits_{l=r+1}^{P}\Gamma(i+n_l+\alpha_l\epsilon)\right)
}{
\left(\prod_{l=1}^{s}\Gamma(i+m_l+\frac{1}{2}+\beta_l\epsilon)\right)
\left(\prod_{l=s+1}^{P-1}\Gamma(i+m_l+\beta_l\epsilon)\right)
}
\frac{x^i}{\Gamma(i+1)}\;,\nonumber\\
\end{eqnarray}
where the integers $n_i$ and $m_i$ can be chosen as appropriate since the choice of the basis function is not unique. The factor $D$ can
be partial-fractioned into a sum of factors $i^n$,
$1/(i+j+\gamma\epsilon)^n$ or $1/(i+j+1/2+\gamma\epsilon)^n$.
\be
D=\sum\limits_{j\ge0,n} \frac{C_{j,n}^+}{(i+j+\gamma\epsilon)^{n}}+\sum\limits_{j<0,n}
\frac{C_{j,n}^+}{(i+j+\gamma\epsilon)^{n}}+\sum\limits_{j,n} \frac{C_{j,n}^{1/2}}{(i+\frac{1}{2}+j+\gamma\epsilon)^{n}}+\sum_n
C_n^-i^n\;.
\ee
Again, we write the terms in $D$ as integration or differentiation operators acting on the basis function
\begin{eqnarray}
\lefteqn{B_s^r(\{n_1,...,n_P;m_1,...,m_{P-1}\},\{\alpha_1,...,\alpha_P;\beta_1,...,\beta_{P-1}\})}&&\nonumber\\
&\equiv&
\frac{
\left(\prod\limits_{l=1}^{s}\Gamma(m_l+\frac{1}{2}+\beta_l\epsilon)\right)
\left(\prod\limits_{l=s+1}^{P-1}\Gamma(m_l+\beta_l\epsilon)\right)
}{
\left(\prod\limits_{l=1}^{r}\Gamma(n_l+\frac{1}{2}+\alpha_l\epsilon)\right)
\left(\prod\limits_{l=r+1}^{P}\Gamma(n_l+\alpha_{l}\epsilon)\right)
}\nonumber\\
&\times&
\sum\limits_{i=1}^\infty\frac{
\left(\prod\limits_{l=1}^{r}\Gamma(i+n_l+\frac{1}{2}+\alpha_l\epsilon)\right)
\left(\prod\limits_{l=r+1}^{P}\Gamma(i+n_l+\alpha_l\epsilon)\right)
}{
\left(\prod_{l=1}^{s}\Gamma(i+m_l+\frac{1}{2}+\beta_l\epsilon)\right)
\left(\prod_{l=s+1}^{P-1}\Gamma(i+m_l+\beta_l\epsilon)\right)
}
\frac{x^i}{\Gamma(i+1)}
\nonumber\\
&=&
{}_PF_{P-1}\left(N_1,...,N_P;M_1,...,M_{P-1};x\right)-1
\end{eqnarray}
obtained by combining the first and last line of Eq.~(\ref{eq:general}). To shorten the notation, we have used
\begin{eqnarray}
N_i=n_i+\frac{1}{2}+\alpha_i\epsilon,\quad 1\le i\le r,&\qquad& N_i=n_i+\alpha_i\epsilon,\quad r<i\le P\,,\nonumber\\
M_i=m_i+\frac{1}{2}+\beta_i\epsilon,\quad 1\le i\le s,&\qquad& M_i=m_i+\beta_i\epsilon,\quad s<i\le P-1\;.\nonumber\\
 \end{eqnarray}
 We see that the function $B$ defined in (\ref{eq:b11}) corresponds to $B_1^1(\{0,1;0\},\{\alpha_1,\alpha_2;\beta_1\})$. The final formula reads
\begin{eqnarray}
\lefteqn{{}_PF_{P-1}\left(A_1,...,A_P;B_1,...,B_{P-1};x\right)}\nonumber\\
&=&
1+\frac{
\left(\prod_{l=1}^{s}\prod\limits_{j_l}^{m_l:b_l}(j_l+\frac{1}{2}+\beta_l\epsilon)\right)
\left(\prod_{l=s+1}^{P-1}\prod\limits_{j_l}^{m_l:b_l}(j_l+\beta_l\epsilon)\right)
}{
\left(\prod\limits_{l=1}^{r}\prod\limits_{j_l}^{n_l:a_l}(j_l+\frac{1}{2}+\alpha_l\epsilon)\right)
\left(\prod\limits_{l=r+1}^{P}\prod\limits_{j_l}^{n_l:a_l}(j_l+\alpha_l\epsilon)\right)
}\nonumber\\
&\times&\left(\sum\limits_{j\ge0,n} C_{j,n}^+J^+(j,n)+\sum\limits_{j<0,n,\gamma} C_{j,n,\gamma}^+J^+(j,n,\gamma)\right.\nonumber\\
&&\qquad\qquad\qquad+\left.\sum\limits_{j,n} C_{j,n}^{1/2}J^+(j+\frac{1}{2},n)+\sum_n C_n^-J^-(0,n)\right)\nonumber\\
&&B_s^r(\{n_1,...,n_P;m_1,...,m_{P-1}\},\{\alpha_1,...,\alpha_P;\beta_1,...,\beta_{P-1}\})\;.\nonumber\\
\end{eqnarray}
This formula relates a hypergeometric function to another one of the same type via differentiation and integration operators. It is
therefore sufficient for the expansion of any ${}_PF_{P-1}$ of type $P_s^r$ to have
\begin{itemize}
\item[a)] the expansion of \emph{one} HF of the type $P_s^r$ (hereafter called the basis HF) 
\item[b)] a procedure to integrate and differentiate the expansion of this basis HF.
\end{itemize}
The expansion of the basis HFs is treated in the next section. The various terms in the expansion will contain HPLs as well as rational
functions. Since the HPLs are constructed by means of iterated integrations,
they are well-suited for carrying out the required integration and differentiation procedures. We will therefore focus on the
derivation of the expansion of the basis functions.
\subsection{All-order expansion of the basis functions}
The achievable depth of expansion of a hypergeometric function with half-integer parameters using our algorithm depends on the available
depth of the expansion of the corresponding basis function $B$. In this section we describe one possible way to obtain an expression
for the basis functions of some types of HF to arbitrary order in the expansion parameter $\epsilon$. 

We first set up some notation. We introduce an associative, distributive but non-commutative product $\otimes$ between HPLs of the same argument
\begin{eqnarray}\label{eq:prodnot}
H(s_1,...,s_n;x)\otimes H(t_1,...,t_m;x)&\equiv& H(s_1,...,s_n,t_1,...,t_m;x)\nonumber\\
\left(H(s_{1...n};x)+H(t_{1...m};x)\right)\otimes H(u_{1...l};x)&\equiv& H(s_{1...n},u_{1...l};x)+H(t_{1...m},u_{1...l};x)\nonumber\\
1\otimes H(t_1,...,t_m;x)&\equiv& H(t_1,...,t_m;x)\;,
\end{eqnarray}
with the short-hand notation $s_{1...n} \equiv s_1,...,s_n$. We can define a ``division'' with respect to this product
\be
H(s_1,...,s_j,...,s_n;x)\oslash H(s_j,...,s_n;x)\equiv H(s_1,...,s_{j-1};x)\;.
\ee
If the vector $t_{1...m}$ is not the last part of the vector  $s_{1...n}$ we define
\be
H(s_{1...n};x)\oslash H(t_{1...m};x)\equiv0\;.
\ee
The advantage of this notation is that the derivative ``factorizes''
\begin{eqnarray}
\frac{d}{d x}\left( H(s_{1...n};x)\otimes H(t_{1...m};x)\right)&=&\left(\frac{d}{dx} H(s_{1...n};x)\right) \otimes H(t_{1...m};x)\nonumber\\
\frac{d^j}{d x^j}\left( H(s_{1...n};x)\otimes H(t_{1...m};x)\right)&=&\left(\frac{d^j}{dx^j} H(s_{1...n};x)\right) \otimes H(t_{1...m};x)\quad j\le n\;.\nonumber\\
\end{eqnarray}
For a $j$-fold derivative applied on a $\otimes$ product whose left HPL has a weight smaller than $j$ we have no factorization, but we
have
\be
\left(\frac{d^j}{d x^j}\left( H(s_{1...n};x)\otimes H(t_{1...m};x)\right)\right)\oslash H(t_{1...m};x) =\left(\frac{d^j}{dx^j}
H(s_{1...n};x)\right)\;.
\ee
\subsubsection{General strategy}\label{sec:generalstrat}
 We make the ansatz
\begin{equation}\label{eq:ansatz}
B=g(x)\left(1+\sum\limits_{j=1}^\infty \epsilon^j\sum\limits_{s_1,...s_j=+,0,-}c(s_1,...,s_j;x)H_{s_1,...,s_j}\left(f(x)\right)\right) \;,
\end{equation}
with $f(x)=\sqrt{x}$ for HFs of type $P^i_{i}$ and
\begin{equation*}
f(x)=i\,\sqrt{\frac{x}{1-x}} \quad {\rm or} \quad f(x)=\frac{1-\sqrt{1-x}}{1+\sqrt{1-x}}
\end{equation*}
for HFs of type $P^i_{i\pm 1}$. These types of arguments have already been used  previously (see for example \cite{Weinzierl:2004bn,Jegerlehner:2002em,Kalmykov:2006pu,Davydychev:2003mv, Kalmykov:2006hu,Kalmykov:2007dk}.) The function $g(x)$ is given by
the value of the HF with the expansion parameter $\epsilon$ put to 0. We see that in the ansatz for HFs of the type $P^i_{i\pm 1}$ we will
get HPLs of imaginary arguments for real arguments of the HF to expand. The properties of HPLs of complex arguments are described in 
Refs.~\cite{Maitre:2007kp,Remiddi,Davydychev:2003mv} as well as in appendix~\ref{sec:analytic}.  

The problem reduces to finding the coefficients $c_{s_1,...,s_j;x}$ in equation (\ref{eq:ansatz}). 
Some of their properties can be stated.
\begin{itemize}
\item The coefficients $c_{s_1,...,s_j;x}$ are homogeneous of order $j$ in the $\alpha_i,\beta_i$, that means that the powers of the
different $\alpha$'s and $\beta$'s sum up to the same power as $\epsilon$. 
\item They have to be symmetric in $\alpha$'s and $\beta$'s corresponding to equal $a$'s or $b$'s.
\item Since an HF ${}_{P+1}F_{P}$ reduces to an HF ${}_PF_{P-1}$ if two of its parameters, one from the first and one from the second
subset, are equal, we have conditions on the
coefficients $c_{s_1,...,s_j;x}$. If one of the $a$'s, say $a_i$, is equal to one of the $b$'s, say $b_j$, the coefficient $c$ should
reduce to the coefficient of the reduced HF when we take the corresponding $\alpha_i$ and $\beta_j$ to be equal.
\end{itemize} 
The hypergeometric function ${}_{P}F_{P-1}$ satisfies a differential equation ${\cal D}$ of order $P$. Inserting the ansatz in the differential equation
\begin{equation*}
{\cal{D}} B =0
\end{equation*}
the left-hand side can also be written as a sum of HPLs with coefficients
\begin{equation} 
\sum\limits_{j=0}^\infty \epsilon^j\sum\limits_{l}\sum\limits_{s_1,...s_l=+,0,-}{\cal D}(s_1,...,s_l)H_{s_1,...,s_l}\left(f(x)\right)=0\;.
\end{equation}
The differential equation is satisfied if all the coefficients ${\cal D}(s_1,...,s_n)$ vanish. We consider the coefficient $\cal D$ of a
given HPL with weight vector $(v_1,...,v_n)$. Since the differential equation is of order $P$, this coefficient will get contributions
from the coefficients $c(...)$ of HPLs of weight $n,n+1,...,n+P$ with weight vectors of the form $(...,v_1,...,v_n)$. Using the notation
(\ref{eq:prodnot}), that is HPLs of the form,
\begin{equation*}
H(s_{1},...,s_{k},f(x))\otimes H(v_{1},...,s_{n},f(x)),\qquad 0\le k\le P.
\end{equation*}
Thus, for the coefficient ${\cal D}(v_{1},...,v_n)$  we only need to consider a part of the ansatz $B$, namely
\begin{eqnarray}
\lefteqn{\tilde B (v_1,...,v_n;x)=c(v_1,...,v_n)g(x)}&&\nonumber\\
&\times&\left(1+\sum\limits_{j=1}^P \epsilon^j\sum\limits_{s_1,...s_j}\tilde c(s_{1...j};x)H(s_{1...j};f(x))\right)\otimes H(v_1,...,v_n,f(x))\;,\nonumber\\
\end{eqnarray}
where we have defined 
\begin{equation}\label{eq:ctilde}
\tilde c(s_{1...j};x)=\frac{c(s_{1...j},v_1,...,v_n;x)}{c(v_1,...,v_n;x)}\;,
\end{equation}
and where the sum over the weights $s_1,...,s_j$ runs over $+,-,0$. The coefficient ${\cal D}(v_1,...,v_n;x)$ is now given by 
\begin{eqnarray}\label{eq:calD}
{\cal D}(v_{1},...,v_{n};x)=\left[\left({\cal{D}}\tilde B (v_{1},...,v_{n};x)\right)\oslash H(v_{1},...,v_{n};x)\right]_{H(...)=0}\;.
\end{eqnarray}
The only dependence of the right hand side on the vector $(v_1,...,v_n)$ is in the overall factor $c(s_{1},...,s_{j};x)$ and, implicitly, in the coefficients $\tilde c(s_{1...j};x)$. 
Our strategy is now to construct coefficients $\tilde c(s_{1...j};x)$ such that the coefficients in Eq.~(\ref{eq:calD}) vanish.
In the following, we illustrate this strategy by means of a simple example.
\subsubsection{Simple example}\label{sec:simpleexample}
We consider the HF 
\be\label{eq:simpleex}
{}_2F_1(\frac{1}{2}+\alpha_1\epsilon,1+\alpha_2 \epsilon;\frac{1}{2}+\beta_1\epsilon;x)\;,
\ee
using the $\pm$ HPL weights defined in \cite{Maitre:2007kp}, we make the ansatz
\be\label{eq:2f1ansatz}
B=\frac{1}{1-x}\left(1+\sum\limits_{j=1}^\infty \epsilon^j\sum\limits_{s_1,...s_j=+,0,-}c(s_1,...,s_j;x)H\left(s_1,...,s_j;\sqrt{x}\right)\right). 
\ee
We have taken $g(x)=(1-x)^{-1}$ since 
\be
{}_2F_1(1,\frac{1}{2};\frac{1}{2};x)=\frac{1}{1-x}\;.
\ee
Since the expansion of Eq.~(\ref{eq:simpleex}) has to be real for negative values of $x$, we have to make sure that the coefficient of
the HPLs  in the expansion (which will have imaginary argument for $x<0$) guarantees that the expansion is real for $x<0$. We know
(see\cite{Maitre:2007kp}) that an HPL of imaginary argument is real if it has an even number of ``+" weights and imaginary if the number
of ``+" weights is odd\footnote{provided the last element of the weight vector is not 0, which is not relevant here since such HPLs are
divergent at $x=0$ and we are only interested in solutions of the differential equation regular at $x=0$.}. Therefore we define
\[\tilde c(s_1,...,s_n;x)=\left\{
\begin{array}{cc}
\sqrt{x}c_o(s_1,...,s_n)    &         \mbox{odd number of + in }s_1,...,s_n\\
c_e(s_1,...,s_n)    &         \mbox{even number of + in }s_1,...,s_n\;.
\end{array}\right.\]
We have now for an even number of ``+'' in $v_{1},...,v_{n}$
\begin{eqnarray}
\lefteqn{\tilde B(v_1,...,v_n;x)=
\frac{c_e(v_1,...,v_n)}{1-x}}&&\nonumber\\  
&\times&\Bigg(1+\epsilon\left(\sqrt{x}c_e(+)H(+;\sqrt{x})+c_e(-)H(-;\sqrt{x})+c_e(0)H(0;\sqrt{x})\right)\nonumber\\ 
&+&\epsilon^2\bigg(c_e(+,+)H(+,+;\sqrt{x})+\sqrt{x}c_e(+,-)H(+,-;\sqrt{x})\nonumber\\ 
&&\qquad+\sqrt{x}c_e(+,0)H(+,0;\sqrt{x})+\sqrt{x}c_e(0,+)H(0,+;\sqrt{x})\nonumber\\ 
&&\qquad+c_e(0,-)H(0,-;\sqrt{x})+c_e(0,0)H(0,0;\sqrt{x})\nonumber\\ 
&&\qquad+\sqrt{x}c_e(-,+)H(-,+;\sqrt{x})+c_e(-,-)H(-,-;\sqrt{x})\nonumber\\
&&+c_e(-,0)H(-,0;\sqrt{x})\bigg)\Bigg)\otimes H(v_1,...,v_n;\sqrt{x})\;, 
\end{eqnarray}
and 
\begin{eqnarray}
\lefteqn{\tilde B(v_1,...,v_n;x)=\frac{c_o(v_1,...,v_n)}{1-x}}&&\nonumber\\  
&\times&\Bigg(\sqrt{x}+\epsilon\left(c_o(+)H(+;\sqrt{x})+c_o(-)\sqrt{x}H(-;\sqrt{x})+c_o(0)\sqrt{x}H(0;\sqrt{x})\right)\nonumber\\ 
&+&\epsilon^2\bigg(c_o(+,+)\sqrt{x}H(+,+;\sqrt{x})+c_o(+,-)H(+,-;\sqrt{x})\nonumber\\
&&\quad+c_o(+,0)H(+,0;\sqrt{x})+c_o(0,+)H(0,+;\sqrt{x})+c_o(0,-)\sqrt{x}H(0,-;\sqrt{x})\nonumber\\
&&\quad+c_o(0,0)\sqrt{x}H(0,0;\sqrt{x})+c_o(-,+)H(-,+;\sqrt{x})\nonumber\\ 
&&\quad+c_o(-,-)\sqrt{x}H(-,-;\sqrt{x})+c_o(-,0)\sqrt{x}H(-,0;\sqrt{x})\bigg)\Bigg)\nonumber\\ 
&\otimes&H(v_1,...,v_n;\sqrt{x}) \;,
\end{eqnarray}
for an odd number of ``+'' in $v_1,...,v_n$. Inserting these expressions into Eq.~(\ref{eq:calD}) and equating the $\epsilon$ coefficient of the left hand side to zero we get conditions on $c_e(+)$, $c_e(-)$, $c_e(0)$, $c_o(+)$ $c_o(-)$  and  $c_o(0)$, namely
\begin{eqnarray}
c_e(+)=\alpha_1-\beta_1 \; ,&\qquad& c_o(+)=\alpha_2\nonumber\\
c_e(-)=\alpha_2 \; ,&\qquad& c_o(-)=\alpha_1-\beta_1\nonumber\\
c_e(0)=0 \; ,&\qquad& c_o(0)=-2\beta_1\;.
\end{eqnarray}
These results are independent of the vector $v_1,...,v_n$, we can thus take this result to define a rule to construct the coefficients
of the HPLs in our ansatz, Eq.~(\ref{eq:2f1ansatz})
\begin{eqnarray}\label{eq:ruleB11}
\lefteqn{c(+,w_1,...,w_n)=}&&\nonumber\\
&&c(w_1,...,w_n)\times\left\{\begin{array}{cc} \alpha_2 &\textnormal{odd number of + in $ \{w_1,...,w_n \}$}\\
\alpha_1-\beta_1&\textnormal{even number of + in $ \{w_1,...,w_n \}$ }
\end{array}\right.\nonumber\\
\lefteqn{c(0,w_1,...,w_n)=}&&\nonumber\\
&&c(w_1,...,w_n)\left\{\begin{array}{cc} -2 \beta_1 &\textnormal{odd number of + in $ \{w_1,...,w_n \}$}\\
0&\textnormal{even number of + in $ \{w_1,...,w_n \}$}
\end{array}\right.\nonumber\\
\lefteqn{c(-,w_1,...,w_n)=}&&\nonumber\\
&&c(w_1,...,w_n)\times\left\{\begin{array}{cc}\alpha_1-\beta_1 &\textnormal{odd number of + in $ \{w_1,...,w_n \}$}\\
\alpha_2 &\textnormal{even number of + in $ \{w_1,...,w_n \}$ }\;.
\end{array}\right.\nonumber\\
\end{eqnarray}
Using these rules we can show that
\be 
{\cal D}(v_1,...,v_n;x)=0,
\ee
for all vectors $(v_1,...,v_n)$. We can ensure that the boundary conditions are respected by setting 
\be
c(0)=0\qquad c(-)=\alpha_1 \qquad c(+)=\alpha_2-\beta_1\;.
\ee
The first terms of the expansion in Eq.~(\ref{eq:simpleex}) are then
\begin{eqnarray}
\lefteqn{{}_2F_1(\frac{1}{2}+\alpha_1\epsilon,1+\alpha_2 \epsilon;\frac{1}{2}+\beta_1\epsilon;x)=}&&\nonumber\\
&&\frac{1}{1-x}\Bigg(1+\epsilon\left(\sqrt{x}(\alpha_1-\beta_1)H(+;\sqrt{x})+\alpha_2 H(-;\sqrt{x})\right)\nonumber\\ 
&+&\epsilon^2\bigg(\alpha_2(\alpha_1-\beta_1)H(+,+;\sqrt{x})+\sqrt{x}(\alpha_1-\beta_1)H(+,-;\sqrt{x})\nonumber\\ 
&&\quad-2\beta_1 (\alpha_1-\beta_1)\sqrt{x}H(0,+;\sqrt{x})\nonumber\\ 
&&\quad+(\alpha_1-\beta_1)^2\sqrt{x}H(-,+;\sqrt{x})+\alpha_2^2H(-,-;\sqrt{x})\bigg)\Bigg)+{\cal O}(\epsilon^3)\;.
 \end{eqnarray}
This expansion could be rewritten in terms of the more usual HPLs of integer weights, yieling the same results as, for example in \cite{Weinzierl:2004bn,Kalmykov:2006hu}, but part of the structure would be lost. The success of the strategy for this case relies on the following facts
\begin{itemize}
\item[a)] The $x$ dependence of the factors $c(...;x)$ must be known and
\item[b)] this dependence must be simple enough in order to minimize the $v_1,...,v_n$ dependence of the $\tilde c(s_{1...j};x)$ in
Eq.~(\ref{eq:ctilde}).
\end{itemize}
\subsection{Applications}
Following this strategy, we found explicit results for the all-order expansions of the following HFs.
\begin{eqnarray*}
&&{}_2F_1(\frac{1}{2}+\alpha_1\epsilon,\alpha_2\epsilon;1+\beta_1\epsilon;x) \qquad {\mbox{of type }} 2^1_0 \; ,\\
&&{}_3F_2(\frac{1}{2}+\alpha_1\epsilon,1+\alpha_2\epsilon,1+\alpha_3\epsilon;\frac{1}{2}+\beta_1\epsilon,1+\beta_2\epsilon;x)\qquad {\mbox{of type }} 3^1_1 \; ,\\
&&{}_3F_2(\alpha_1\epsilon,1+\alpha_2\epsilon,1+\alpha_3\epsilon;\frac{1}{2}+\beta_1\epsilon,1+\beta_2\epsilon;x)\qquad {\mbox{of type
}} 3^0_1\; , \\
&&{}_3F_2(\frac{1}{2}+\alpha_1\epsilon,\frac{1}{2}+\alpha_2\epsilon,1+\alpha_3\epsilon;\frac{1}{2}+\beta_1\epsilon,\frac{1}{2}+\beta_2\epsilon;x)\qquad
{\mbox{of type }} 3^2_2 \; .
\end{eqnarray*}
They are displayed in appendix \ref{sec:basis}. Some of the expansions of those types listed at the
beginning of the next section are too lengthy to be reported here. The result of the expansions found in all the expansions are expressed through HPLs of integer and $\pm$ weights of arguments 
\[\sqrt{x},\quad i\sqrt{\frac{x}{1-x}},\quad \frac{1-\sqrt{1-x}}{1+\sqrt{1-x}}.\]
The analytic continuation of the expansion for these arguments is addressed in Appendix \ref{sec:analytic}.
  
In section \ref{sec:basis312} we show that the all order expansion of 
\begin{equation*}
{}_3F_2\left(\frac{1}{2}+\alpha_1\epsilon,\alpha_2\epsilon,1+\alpha_3\epsilon;\frac{1}{2}+\beta_1\epsilon,\frac{1}{2}+\beta_2\epsilon;x\right)
\qquad {\mbox{of type }} 3^1_2\\
\end{equation*}
cannot be expressed within the basis of HPLs used in this work.    
\section{Extension of the Mathematica package {\tt HypExp}}\label{manual}
%
We used the algorithm of the preceding section to extend the capabilities of the existing {\tt Mathematica} package {\tt HypExp} \cite{Huber:2005yg}. The
second version of the package is now able to expand, in addition to HFs with only integer parameters, $_JF_{J-1}$-functions of the
types
\[2^2_1,\quad 2^1_1,\quad 2^1_0,\quad 2^0_1,\]
\[3^3_2,\quad 3^2_2,\quad 3^1_1,\quad 3^1_0,\quad 3^0_1,\]
\[ 4_1^1,\quad 4_3^3\]
to arbitrary order in a small quantity about their parameters. The limitation to the number of classes of functions that can be expanded is due to the fact that the all-order expansion of a basis function for each class has to be computed on a cases-by-case basis. The package is publicly available from~\cite{homepage}. The results are displayed in terms of rational functions,
logarithms, polylogarithms, Nielsen polylogarithms, and HPLs. To treat the harmonic polylogarithms, the package {\tt HypExp} uses the
package {\tt HPL}~\cite{Maitre:2005uu} whose second version \cite{Maitre:2007kp} implements the harmonic polylogarithms of complex
arguments, as required by the expansion about half-integer parameters. The results given by the package are valid in the entire complex
plane. However, along the branch cut running along the real axis from 1 to $+\infty$ care has to be taken in order to reproduce the
correct imaginary part, see section~\ref{sec:examples} as well as appendix~\ref{sec:analytic}.

In this section we shortly describe the new features of the package {\tt HypExp}, for a more detailed description, we refer to our previous work \cite{Huber:2005yg}.

After installation\footnote{see link~\cite{homepage} for more information on the installation procedure}, the package {\tt HypExp} may be loaded using the following command,
\vspace{0.2cm}\\
\fbox{\parbox{0.75\textwidth}{
\example
}
}\vspace{0.2cm}\\
This should be done at the beginning of the session. 

{\tt HypExp[Hypergeometric2F1[\dots,x],$\epsilon$,n]} or \\ {\tt HypExp[HypergeometricPFQ[\dots,x],$\epsilon$,n]} returns the
$\epsilon$-expansion of the enclosed hypergeometric function (HF) through order ${\cal O}(\epsilon^n)$ if the type of HF is supported.
\vspace{0.2cm}\\
\fbox{\parbox{\exboxlength}{
\example\\
\example\\
\example\\
\example
}
}
 
The function {\tt HypExp} applied to anything else but a HF will return the same object.
The result is not returned as a {\tt SeriesData} object because this would have the effect of forcing the expansion of the rest of the
expression.

The prefactors that accompany the variable $\dps \eps$ can also be symbolic, and the expansion also works for argument $x=1$ as shown by
the following examples,
\vspace{0.2cm}\\
\fbox{\parbox{\exboxlength}{
\example\\
\example\\
\example\\
\example
}
}\vspace{0.2cm}\\
The package also updates {\tt Series} to allow it to expand compound expressions containing hypergeometric functions. The difference
between this and the function {\tt HypExp} is that the other functions of $\epsilon$ are also expanded.
\vspace{0.2cm}\\
\fbox{\parbox{1.05\textwidth}{
\example\\
\example\\
\example\\
\example
}
}\vspace{0.2cm}\\
The use of the function {\tt HypExp} is preferable in particular if the expansion order is high or the expression containing the HFs is
large. The results given by the package are not systematically simplified using {\tt Simplify}, since the simplification might take longer than
the expansion itself, in particular for expansions to high orders. A {\tt Simplify} might produce a more compact result.
For additional commands and functions implemented in the package see~\cite{Huber:2005yg}.
The package was developed in {\tt Mathematica 5} and tested on both versions 5 and 6. 

\section{Examples and applications}\label{sec:examples}

A first application of the present package has already been given in Ref.~\cite{Grozin:2007ap} for the case of a three-loop master
integral with HQET propagators. Four more examples with up to four loops are given below.

\subsection{One-loop vertex correction}

\begin{figure}[t]
\begin{picture}(340,120)
\LongArrow(65,60)(45,60)
\Line(45,60)(25,60)
\Line(65,60)(105,100)
\Line(65,60)(105,20)
\DashLine(105,20)(105,100){5}
\LongArrow(145,100)(125,100)
\Line(125,100)(105,100)
\LongArrow(145,20)(125,20)
\Line(125,20)(105,20)
\Text(128,110)[]{$p_1^2=m^2$}
\Text(128,10)[]{$p_2^2=m^2$}
\Text(35,70)[]{$s = (p_1+p_2)^2$}
\Text(85,90)[]{$m$}
\Text(85,30)[]{$m$}
\Line(230,60)(250,60)
\LongArrow(210,60)(230,60)
\DashLine(250,60)(290,100){5}
\DashLine(250,60)(290,20){5}
\DashLine(290,20)(290,100){5}
\DashLine(250,60)(273,60){4}
\DashLine(273,60)(290,100){5}
\DashLine(273,60)(290,20){5}
\DashArrowLine(290,100)(330,100){5}
\DashArrowLine(290,20)(330,20){5}
\Text(313,110)[]{$p_1^2=0$}
\Text(313,10)[]{$p_2^2=0$}
\Text(220,70)[]{$q^2 = (p_1+p_2)^2$}
\end{picture}
\caption{Left panel: One-loop vertex correction with one massless line (dashed) and two massive ones (solid) with equal masses. Right
panel: Three-loop master integral with six internal massless lines.
}\label{fig:examplediagrams1}
\end{figure}
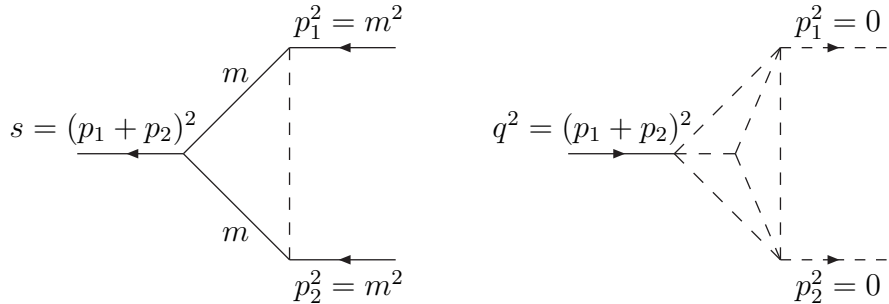

Our first example is the one-loop vertex correction displayed on the left in Fig.~\ref{fig:examplediagrams1}, of which a one-dimensional
Mellin-Barnes representation was derived in Ref.~\cite{Gluza:2007rt}. Assuming all propagators
to be raised to unit power, we get
\begin{eqnarray}
  I &\equiv& \dps \int \! \left[dk\right] \frac{1}{\left[k^2-m^2\right]\left[(k-p_1-p_2)^2-m^2\right]\left[(k-p_1)^2\right]}\nonumber\\
  &=&  \dps \frac{e^{\epsilon\gamma_E}}{2\epsilon}\MB{}{z} \, \frac{\Gamma(-z) \, \Gamma^2(-\epsilon-z) \,
  \Gamma(1+\epsilon+z)}{\Gamma(-2\epsilon-2z)} \, \left[m^2\right]^z \, \left[-s-i\eta\right]^{-1-\epsilon-z}
\label{eq:ex1loopmb}
\end{eqnarray}
with
\be\label{eq:measure}
\dps \left[dk\right]=\dps \frac{e^{\epsilon \gamma_E}}{i \pi^{D/2}} \, d^Dk
\ee
and $D=4-2\epsilon$. Here and in the following we tacitly assume that all propagators contain an infinitesimal $+i\eta$. The integration
contour in Eq.~(\ref{eq:ex1loopmb}) is chosen such that it separates left poles of the
$\Gamma$-functions from right ones. Applying the formula
\begin{equation}\label{eq:gammadouble}
\Gamma(2z) = \frac{2^{2z-1}}{\sqrt{\pi}} \, \Gamma(z) \, \Gamma(z+\frac{1}{2})
\end{equation}
to the $\Gamma$-function in the denominator of Eq.~(\ref{eq:ex1loopmb}) and subsequently summing the residues to the left of the
contour, we obtain
\be
 I =  \dps \frac{e^{\epsilon\gamma_E}}{2\epsilon} \, \Gamma(1+\epsilon) \left[m^2\right]^{-1-\epsilon}
 \pFq{2}{1}{1,1+\epsilon}{\frac{3}{2}}{\hat s} \, ,
\label{eq:ex1loopclosed}
\ee
with $\hat s \equiv (s+i\eta)/(4m^2)$. We demonstrate below how the $\epsilon$-expansion of this expression can be performed with the
package (we set $m=1$).
\vspace{0.2cm}\\
\fbox{\parbox{0.95\textwidth}{\includegraphics[scale=.79]{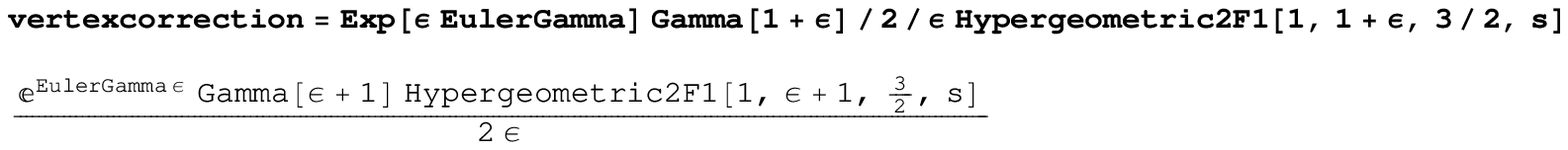}}
}\vspace{0.2cm}\\
\vspace{0.2cm}\\
\fbox{\parbox{0.95\textwidth}{\includegraphics[scale=.79]{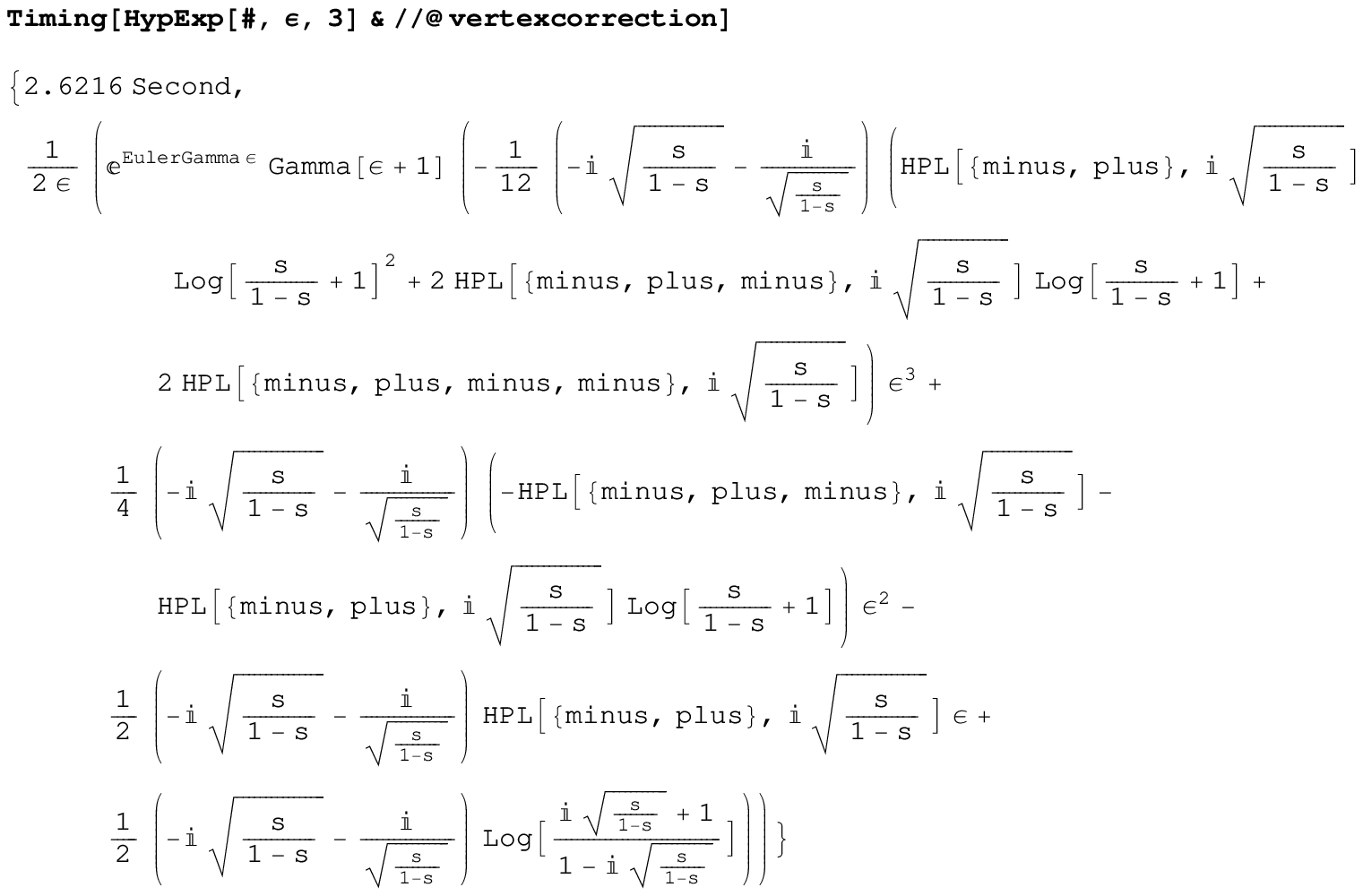}}
}\vspace{0.2cm}
\vspace{0.2cm}\\
\fbox{\parbox{0.95\textwidth}{\includegraphics[scale=.79]{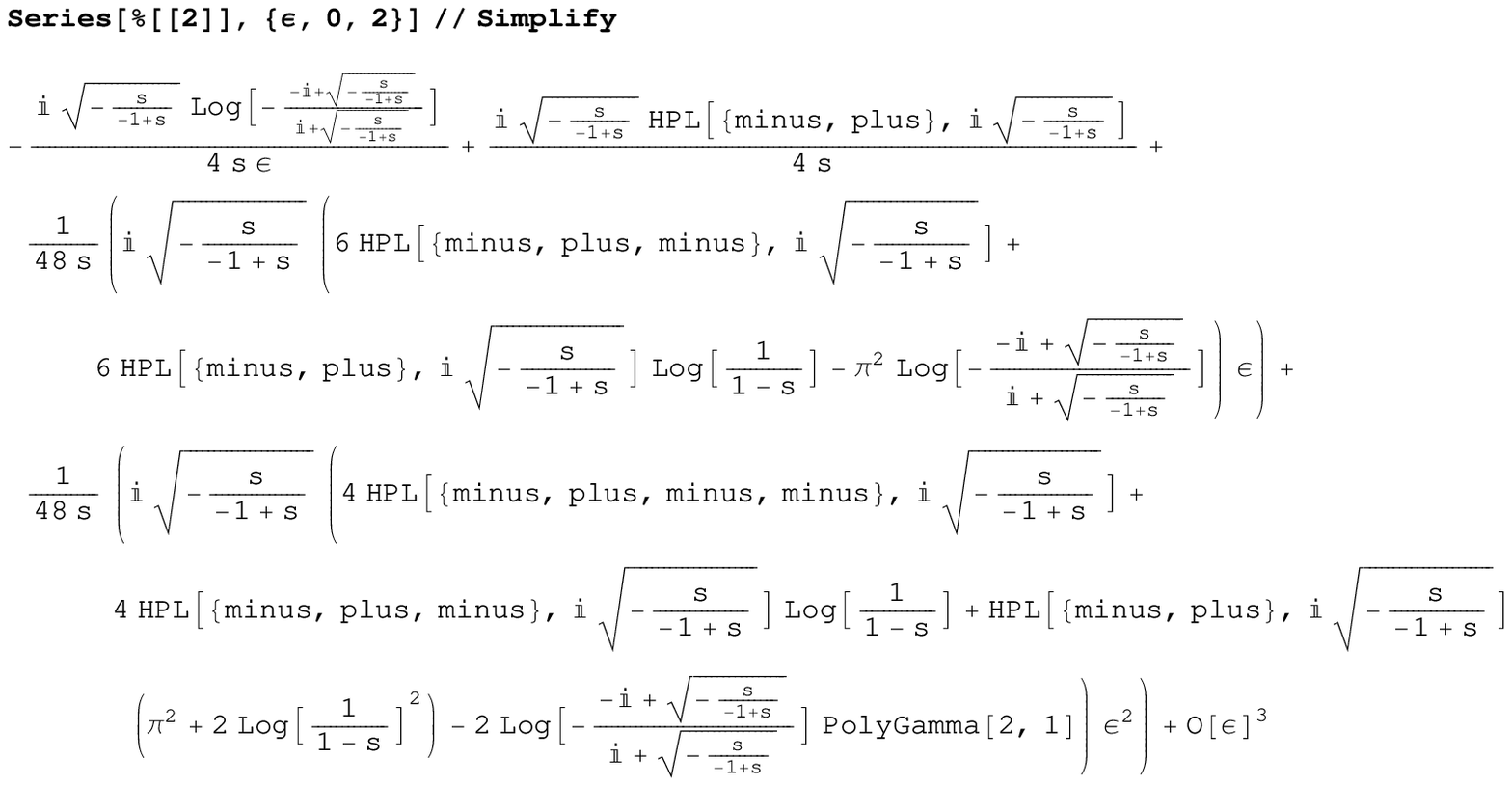}}
}\vspace{0.2cm}

The result of the expansion agrees with the one recently given in Ref.~\cite{Gluza:2007bd}. By means of the commands {\tt HPLpm21m1} and
{\tt HPLConvertToKnownFunctions} from the {\tt HPL} package~\cite{Maitre:2005uu,Maitre:2007kp}, the above expansion can be converted to
more common functions like logarithms, polylogarithms, and Nielsen polylogarithms. As a cautionary remark, we would like to stress that
above threshold the sign of the $i\eta$-prescription in Eq.~(\ref{eq:ex1loopclosed}) is crucial in order to obtain the correct result,
see appendix~\ref{sec:analytic} and Refs.~\cite{Maitre:2005uu,Maitre:2007kp}.

\subsection{Massless three-loop master integral}
Our second example is the computation of the massless three-loop master integral which was identified in Ref.~\cite{Gehrmann:2006wg} and
which is depicted on the right in Fig.~\ref{fig:examplediagrams1}.
After Feynman parametrization, this integral reads
\begin{eqnarray}
\dps A_{6,2} &=& {\int \!\!\! \frac{d^D k}{\left(2\pi\right)^D}\!} {\int \!\!\! \frac{d^D l}{\left(2\pi\right)^D}\!} {\int \!\!\!
\frac{d^D r}{\left(2\pi\right)^D}\!} \; \; \frac{1}{\left( k+p_1\right)^2 \, \left( k+l-p_2\right)^2 } \nnb \\
&\times& \frac{1}{\left. l \right.^2 \, \left. r \right.^2 \, \left(r-k\right)^2 \, \left( r-k-l\right)^2}\\
&=&-{\cal N} \, \Gamma^3(1-\epsilon )\,\Gamma (3 \epsilon )\int\limits_0^1 du\,dt\,dz\,dy\,dr\,   \nonumber\\
&\times& D^{-3\epsilon } (1-r)^{-3 \epsilon }  (1-t)^{-\epsilon } (1-y)^{\epsilon }(1-z)^{1-2 \epsilon }
    r^{-\epsilon } z^{-3 \epsilon } \nonumber\\
&\times& \bigg(1+r \Big((1-z) z D^2+(1-y) (1-z) t D\nonumber\\
&&\qquad+(1-u) (1-y)^2 (1-z) t^2 u-1\Big)\bigg)^{4\epsilon -2}
\end{eqnarray}
with
\begin{equation}
D=1-(1-y)(1-(1-t) (1-u) u) ,\quad \mbox{and}\quad 
{\cal N}=\frac{i \, (4\pi)^{3 \epsilon -6}}{ \Gamma^3(1-\epsilon )}\left(-q^2-i\eta\right)^{-3 \epsilon } \, .
\end{equation}
After numerous variable changes we get
\begin{eqnarray}
A_{6,2}&=&\frac{{\cal N} \, 2^{8 \epsilon -2}\, \pi \,  \Gamma^2(1-3 \epsilon )
   \Gamma^5(1-\epsilon ) \Gamma (3 \epsilon )
  }{\epsilon  \, \Gamma (2-4 \epsilon ) \Gamma^2\!\left(\frac{3}{2}-2 \epsilon \right)}\nonumber\\
&\times&\int\limits_0^1 \! ds \, (1-s)^{-3\epsilon } s^{2 \epsilon -1} \left(\frac{\Gamma (1-\epsilon )^2}{\Gamma (1-2
   \epsilon )}-s^{-\epsilon } \, _2F_1(\epsilon
   ,-\epsilon ;1-\epsilon ;s)\right) \nonumber\\
&\times&  _3F_2\left(1-3
   \epsilon ,1-2 \epsilon ,1-\epsilon ;2-4 \epsilon
   ,\frac{3}{2}-2 \epsilon ;-\frac{(s-1)^2}{4
   s}\right) \, .
\end{eqnarray}
We see that we have an HF of type $3_1^0$ with one half integer parameter. We can expand each factor in $\epsilon$. The powers of $s$
and $(1-s)$ can be expressed using HPLs (and HPL weight functions)
\begin{eqnarray}
(1-s)^{-3 \epsilon }&=&\sum_j 3^j\epsilon^j H({}^j1;s)\nonumber\\
&=& 1+3\epsilon H(1;s) +9\epsilon ^2 H(1,1;s)+27 \epsilon ^3  H(1,1,1;s)+\dots  \nonumber\\
(s)^{-\epsilon }&=&\sum_j (-1)^j\epsilon^j H({}^j1;s)\nonumber\\
&=& 1+2\epsilon H(0;s) +4\epsilon ^2 H(0,0;s)+8 \epsilon ^3  H(0,0,0;s)+\dots  \nonumber\\
(s)^{-1+2\epsilon }&=&\frac{1}{s}\left(\sum_j 2^j\epsilon^j H({}^j1;s)\right)\nonumber\\
&=& \frac{1}{s}\left(1+2\epsilon H(0;s) +4\epsilon ^2 H(0,0;s)+8 \epsilon ^3  H(0,0,0;s)+\dots  \right)\; .\nonumber\\
\end{eqnarray}
The HFs can be expanded with {\tt HypExp}. The first one includes only integer parameters and hence its expansion will contain only HPLs
of argument $s$. The second one has half-integer parameters and will have HPLs of square root arguments
\[i\,\sqrt{\frac{x}{1-x}},\qquad\quad \mbox{with}\qquad\quad x=-\frac{(1-s)^2}{4s}\, .\] 
Since $s$ is in the interval $(0,1)$, we have
\[i\,\sqrt{\frac{x}{1-x}}=-\frac{1-s}{1+s}\;,\]
so that we can convert the HPLs of this argument into HPLs of argument $s$ by applying (twice) the command {\tt
HPLConvertToSimplerArguments} from the {\tt HPL} package. The next step is to expand the product of HPLs into a sum of
HPLs which we can then integrate by means of the integration routines of {\tt HPL}. This procedure
is not restricted to a specific depth of the expansion, so we could, in principle, expand $A_{6,2}$ to all orders. After factoring out
an appropriate combination of prefactors, the result can be written as the following homogeneous sum, where the power of $\epsilon$
grows with the transcendentality of the coefficient.
\begin{eqnarray}
A_{6,2}&=&\frac{{\cal N}}{(1-5 \epsilon ) (1-4 \epsilon )
   \epsilon }\nonumber\\
&\times&\bigg[-2 \zeta (3)-\epsilon\,\frac{7 \pi ^4}{180}+ \epsilon ^2\left(\frac{2}{3} \pi ^2 \zeta (3)-10 \zeta(5)\right)\nonumber\\
   &&+\epsilon ^3\left(\frac{163 \pi
   ^6}{7560}+76 \zeta (3)^2\right) +\epsilon^4\left(\frac{55}{18} \pi ^4 \zeta (3)+\frac{445
   \zeta (7)}{2}\right) \nonumber\\
   &&+ \epsilon^5\left(-\frac{744}{5} \zeta(5,3)+1000
   \zeta (3) \zeta (5)-22 \pi ^2 \zeta
   (3)^2+\frac{802183\pi ^8}{4536000}\right)+{\cal O}(\epsilon^6)\bigg]\, ,\nonumber\\ 
\end{eqnarray}
where we have encountered a multiple zeta value in the last term.

\subsection{Two-loop vacuum diagram with three massive lines of equal masses}

\begin{figure}[t]
\begin{picture}(340,120)
\CArc(60,60)(60,0,360)
\Line(60,0)(60,120)
\CArc(260,60)(60,0,360)
\Line(260,0)(260,120)
\DashCArc(320,60)(84.85,135,225){5}
\DashCArc(200,60)(84.85,315,45){5}
\end{picture}
\caption{Left panel: Two-loop vacuum diagram with three massive lines of equal masses. Right panel: Four-loop tadpole diagram with two
massless lines (dashed) and three massive ones (solid) with equal masses.}\label{fig:examplediagrams2}
\end{figure}
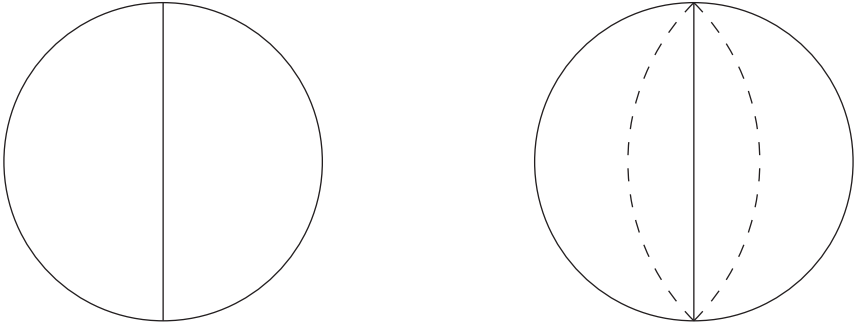

Our third expample is the two-loop vacuum diagram with three massive lines of equal masses which is displayed on the left in
Fig.~\ref{fig:examplediagrams2}. It has been considered at several places in the literature, see e.g.\
Refs.~\cite{Davydychev:1992mt,Chetyrkin:1997fm,Schroder:2005va,Davydychev:1995mq,Fleischer:1999tu,Davydychev:1999mq,Davydychev:2000na}. In Ref.~\cite{Davydychev:1992mt} the following expression for positive
integer values of $n_1$, $n_2$, and $n_3$ was derived in terms of hypergeometric functions,
\begin{eqnarray}
I_{n_1 n_2 n_3} &\equiv& \int \!\! d^D k \int \!\! d^D l \; \;
\frac{1}{\left[k^2-m^2\right]^{n_1}\left[l^2-m^2\right]^{n_2}\left[(k-l)^2-m^2\right]^{n_3}}\nonumber\\
&=&\pi^D \, (-1)^{1-n_{123}} \, (m^2)^{D-n_{123}}\nonumber\\
&\times&
\left\{\frac{\Gamma(D/2-n_3)\,\Gamma(n_{13}-D/2)\,\Gamma(n_{23}-D/2)\,\Gamma(n_{123}-D)}{\Gamma(D/2)
\,\Gamma(n_1)\,\Gamma(n_2)\,\Gamma(n_{1233}-D)}\right.\nonumber\\
&&\times \, {}_4 F_3\left( \left.
\begin{array}{c}
n_3,n_{13}-D/2,n_{23}-D/2,n_{123}-D\\
(n_{1233}-D)/2,(n_{1233}-D+1)/2,n_3-D/2+1
\end{array}\right|\frac{1}{4}\right)\nonumber\\
&& + \frac{\Gamma(n_3-D/2)\,\Gamma(n_{12}-D/2)}{\Gamma(n_3)\,\Gamma(n_{12})}\nonumber\\
&&\times \left.\, {}_4 F_3\left( \left.
\begin{array}{c}
n_1,n_2,n_{12}-D/2,D/2\\
(n_{12})/2,(n_{12}+1)/2,D/2-n_3+1
\end{array}\right|\frac{1}{4}\right)\right\}\; ,\label{eq:davytausk}
\end{eqnarray}
with the usual abbreviation
\be
 n_{1\ldots j} = n_1+\ldots+n_j \, .
\ee
It turns out that for positive
integer values of $n_1$, $n_2$, and $n_3$ the ${}_4 F_3$ functions in Eq.~(\ref{eq:davytausk}) can be expressed as a linear combination
of ${}_2 F_1$ functions of type $2_1^0$, which can then be expanded by means of the package. For instance, for $n_1=n_2=n_3=1$ we obtain
the known result~\cite{Davydychev:1992mt,Chetyrkin:1997fm,Schroder:2005va,Davydychev:1995mq,Fleischer:1999tu,Davydychev:1999mq,Davydychev:2000na}
\begin{eqnarray}
I_{111} &=& \pi^{4-2\epsilon} \, (m^2)^{1-2\epsilon} \, A(\epsilon) \,
\left\{\frac{3}{2\epsilon^2}-\frac{2 \pi^2}{3}+\psi^{(1)}\!\!\left(\textstyle{\frac{1}{3}}\displaystyle\right)\right.\nonumber\\
&&\left.+\epsilon \left(\frac{3}{5}\,\Phi\!\left(\textstyle{-\frac{1}{3}}\displaystyle,3,\textstyle{\frac{1}{2}}\displaystyle\right) -
\frac{17\pi^3}{30\sqrt{3}}-\frac{\pi\sqrt{3}}{10} \, \ln^2(3)\right)+{\cal O}(\epsilon^2)\right\} \; ,\label{eq:I111}
\end{eqnarray}
with
\begin{equation}
A(\epsilon) = \frac{\Gamma^2(1+\epsilon)}{(1-\epsilon)(1-2\epsilon)} \; .
\end{equation}
In Eq.~(\ref{eq:I111}) we have used a slightly different notation compared to the results given previously in the
literature~\cite{Davydychev:1992mt,Chetyrkin:1997fm,Schroder:2005va,Davydychev:1995mq,Fleischer:1999tu,Davydychev:1999mq}. Relations among the different notations involve (see also
Refs.~\cite{Broadhurst:1987ei,Davydychev:2000na})
\begin{eqnarray}
 \psi^{(1)}\!\!\left(\textstyle{\frac{1}{3}}\displaystyle\right) &=& \frac{2\pi^2}{3} + \frac{27}{2} \, S_2 = \frac{2\pi^2}{3} + 3\sqrt{3}
 \, {\rm Ls}_2(\textstyle{\frac{2\pi}{3}}\displaystyle) \nonumber\\
 &=& \frac{2\pi^2}{3} +\frac{\pi\sqrt{3}}{5}\, \ln(3) +
 \frac{3}{5}\,\Phi\!\left(\textstyle{-\frac{1}{3}}\displaystyle,2,\textstyle{\frac{1}{2}}\displaystyle\right)\\
 {\rm Ls}_3(\textstyle{\frac{2\pi}{3}}\displaystyle) &=& -\frac{16\pi^3}{135}-\frac{2\pi^2}{9\sqrt{3}}\,\ln(3)
 -\frac{\pi}{30}\,\ln^2(3)\nonumber\\
 && + \frac{\ln(3)}{3\sqrt{3}}\,\psi^{(1)}\!\!\left(\textstyle{\frac{1}{3}}\displaystyle\right)+\frac{1}{5\sqrt{3}}\,
 \Phi\!\left(\textstyle{-\frac{1}{3}}\displaystyle,3,\textstyle{\frac{1}{2}}\displaystyle\right)\\
 S_2 &=& -\frac{4}{9\sqrt{3}} \int_0^{\pi/3} dx \, \ln|2\sin(\textstyle{\frac{x}{2}}\displaystyle)|\label{eq:s2def}\\
 {\rm Ls}_j(\theta) &=& -\int_0^{\theta} dx \, \ln^{j-1}|2\sin(\textstyle{\frac{x}{2}}\displaystyle)|\\
 \Phi\!\left(z,s,a\right) &=& \sum\limits_{k=0}^{\infty} \frac{z^k}{[(k+a)^2]^{s/2}} \; .
\end{eqnarray}

\subsection{Four-loop tadpole with three massive lines}

Our last example is related to the previous one, namely the four-loop tadpole with three equal massive lines displayed on the right in
Fig.~\ref{fig:examplediagrams2}. We follow again Ref.~\cite{Gluza:2007rt}, where the following one-dimensional Mellin-Barnes
representation was derived
\begin{eqnarray}
  T(n_1,n_2,n_3,n_4,n_5) &\equiv& \dps \int \! \left[dk_1\right] \! \int \! \left[dk_2\right] \! \int \! \left[dk_3\right]\! \int \!
  \left[dk_4\right] \frac{1}{\left[k_1^2-m^2\right]^{n_1}}\nonumber\\
  &&\hspace*{-30pt}\dps\times \frac{1}{\left[(k_1+k_2)^2-m^2\right]^{n_2}\left[(k_2+k_3+k_4)^2-m^2\right]^{n_3}\left[k_4^2\right]^{n_4}
  \left[k_3^2\right]^{n_5}}\nonumber\\
  &=&  \dps \frac{(-1)^{n_{12345}} \, (m^2)^{8-4 \epsilon-n_{12345}} \,
  e^{4\epsilon\gamma_E}}{\Gamma(n_1)\,\Gamma(n_2)\,\Gamma(n_3)\,\Gamma(n_4)\,\Gamma(n_5)}\nonumber\\
&\times& \frac{\Gamma(2- \epsilon-n_4) \Gamma(2-\epsilon-n_5)}
       {\Gamma(2- \epsilon)}
  \MB{}{z} \, \Gamma(2-\epsilon-n_1-z)
\nonumber\\
&\times& \Gamma(-z)
         \Gamma(2-\epsilon-n_2-z)
         \Gamma(4-2\epsilon-n_{12}-z)
\nonumber\\
&\times& \frac{
         \Gamma(-6+3\epsilon+n_{1245}+z)
         \Gamma(-8+4\epsilon+n_{12345}+z)}
         {\Gamma(4-2\epsilon-n_{12}-2z)} ,
\label{eq:ex1mb}
\end{eqnarray}
with the same integration measure as in Eq.~(\ref{eq:measure}).
We shall again restrict ourselves to the case in which all the
$n_i$ as well as the mass are equal to unity. We apply again formula~(\ref{eq:gammadouble})
and obtain
\begin{eqnarray}
  T(1,1,1,1,1) &=& -\frac{e^{4\epsilon\gamma_E} \, \Gamma^2(1-\epsilon)}{\Gamma(2-\epsilon)} \, 2^{2\epsilon-1} \, \sqrt{\pi} 
  \nonumber\\
&\times& \MB{}{z} \, \Gamma(-z) \Gamma(2-z-2\epsilon)  \nonumber\\
&\times& \frac{\Gamma(1-z-\epsilon) \Gamma(z+3\epsilon-2) \Gamma(z+4\epsilon-3)}{\Gamma(\frac{3}{2}-z-\epsilon)} \,
\left(\frac{1}{4}\right)^{-z}\,.
 \label{eq:ex1nisone}
\end{eqnarray} 
This expression can now be converted to a Meijer-G function~\cite{thebook,thebook2,Meijer1,Meijer2,Meijer3} and subsequently be
expanded in terms of hypergeometric functions. This procedure is in this case equivalent to summing all residues of left poles of
$\Gamma$-functions in Eq.~(\ref{eq:ex1nisone}). The result can be displayed in the following closed form
\begin{eqnarray}
  T(1,1,1,1,1) &=& \frac{2^{3-4\epsilon} \, e^{4\epsilon\gamma_E} \, \pi \, \Gamma^2(1-\epsilon)}{\sin(\pi \epsilon) \,
  \Gamma(2-\epsilon)}\left[\frac{\sqrt{\pi} \, \Gamma(\epsilon) \Gamma(-1+2\epsilon) \Gamma(-2+3\epsilon)}{\Gamma(2-\epsilon)
  \Gamma(-\frac{1}{2}+2\epsilon)}\right.\nonumber\\
&&\times
\pFq{3}{2}{\epsilon,-1+2\epsilon,-2+3\epsilon}{2-\epsilon,-\textstyle\frac{1}{2}\displaystyle+2\epsilon}{\textstyle
\frac{1}{4}\displaystyle}\nonumber\\
&&-\frac{\Gamma(-\frac{1}{2}+\epsilon)\Gamma(-2+3\epsilon)\Gamma(-3+4\epsilon)}{\Gamma(-\frac{3}{2}+3\epsilon)}\nonumber\\
&&\times
\pFq{3}{2}{-1+2\epsilon,-2+3\epsilon,-3+4\epsilon}{\epsilon,-\textstyle\frac{3}{2}\displaystyle+3\epsilon}{\textstyle
\frac{1}{4}\displaystyle}\bigg] \, ,
 \label{eq:ex1closed}
\end{eqnarray}
which is well-suited for expansion with the {\tt HypExp} package. After some manipulations and simplifications on the harmonic
polylogarithms, one gets for the expansion in $\epsilon$ up to the finite part
\begin{eqnarray}
  T(1,1,1,1,1) &=& \frac{1}{4\epsilon^4} + \frac{1}{\epsilon^3}+\left(\frac{97}{48}+\frac{\pi^2}{12}\right)\frac{1}{\epsilon^2}
                    + \left(\frac{833}{288}+\frac{\pi^2}{3}-\frac{\zeta_3}{3}\right)\frac{1}{\epsilon}+\frac{4177}{432}
  \nonumber\\
&&\hs{-20}+\frac{97\pi^2}{144}-\frac{4\zeta_3}{3}+\frac{\pi^4}{12}+\frac{1}{1728}\left[99+16\pi^2-24\,
\psi^{(1)}\!\!\left(\textstyle{\frac{1}{3}}\displaystyle\right)\right]^2+{\cal O}(\epsilon)\, ,\nonumber\\
 \label{eq:ex1expanded}
\end{eqnarray}
in agreement with the findings of Refs.~\cite{Boughezal:2006xk,Gluza:2007rt,Faisst:2006sr}\footnote{Note that $s_2$ in
Ref.~\cite{Faisst:2006sr} is defined differently from $S_2$ in Eq.~(\ref{eq:s2def})}. As a closing remark, we mention that hypergeometric functions of argument
$z=1/4$ as in our last two examples have been studied extensively in the literature and the structure of the coefficients in their expansion has been
analysed~\cite{Broadhurst:1998rz,Fleischer:1999mp,Davydychev:2000na}.
\section{Conclusion and outlook}\label{sec:conclusions}
In this paper we have presented a new algorithm to expand hypergeometric functions about half-integer parameters. The strategy is to
express a HF of a given type in terms of integration and differentiation operators acting on a particular HF of the same type, the
latter is called the basis function. The choice of the basis function for a given type of HF is not unique and can be taken such as to
be as convenient as possible. The method therefore requires as ingredients the knowledge of
\begin{itemize} 
\item[a)] The expansion of \emph{one} HF of the same type as the one to be expanded, namely the basis function.
\item[b)] Integration and differentiation routines for the functions occurring in this expansion. 
\end{itemize}  
The method is quite general and might be applied to HFs ${}_PF_{P-1}$ of higher $P$ or to other types of HFs.

We explain a strategy to compute explicit all-order expansions of the hypergeometric functions and give explicit results for

\vspace*{-10pt}

\begin{eqnarray*}
&&{}_2F_1(\frac{1}{2}+\alpha_1\epsilon,\alpha_2\epsilon;1+\beta_1\epsilon;x) \qquad {\mbox{of type }} 2^1_0 \; ,\\
&&{}_3F_2(\frac{1}{2}+\alpha_1\epsilon,1+\alpha_2\epsilon,1+\alpha_3\epsilon;\frac{1}{2}+\beta_1\epsilon,1+\beta_2\epsilon;x) \qquad {\mbox{of type }} 3^1_1 \; ,\\
&&{}_3F_2(\alpha_1\epsilon,1+\alpha_2\epsilon,1+\alpha_3\epsilon;\frac{1}{2}+\beta_1\epsilon,1+\beta_2\epsilon;x) \qquad {\mbox{of type }} 3^0_1 \; ,\\
&&{}_3F_2(\frac{1}{2}+\alpha_1\epsilon,\frac{1}{2}+\alpha_2\epsilon,1+\alpha_3\epsilon;\frac{1}{2}+\beta_1\epsilon,\frac{1}{2}+\beta_2\epsilon;x) \qquad {\mbox{of type }} 3^2_2 \; .\\
\end{eqnarray*}

\vspace*{-20pt}

By taking appropriate limits, these expansions yield the expansion of the HFs
\begin{eqnarray*}
&&{}_2F_1(\frac{1}{2}+\alpha_1\epsilon,1+\alpha_2\epsilon;\frac{1}{2}+\beta_1\epsilon;x)\\
&&{}_2F_1(1+\alpha_1\epsilon,1+\alpha_2\epsilon;\frac{1}{2}+\beta_1\epsilon;x)\;.
\end{eqnarray*} 
The algorithm described here has been implemented for HFs of the types
\[2^2_1,\quad 2^1_1,\quad 2^1_0,\quad 2^0_1,\]
\[3^3_2,\quad 3^2_2,\quad 3^1_1,\quad 3^1_0,\quad 3^0_1,\] 
\[ 4_1^1,\quad 4_3^3\]
in the already existing Mathematica Package {\tt HypExp}.

In order to show the relevance of hypergeometric functions with half-integer parameters
and to demonstrate the performance of the package we have given examples of Feynman diagrams with up to four loops. We reproduced known
results, but also gave new results as for instance in the case of the massless three-loop master integral $A_{6,2}$.
\section*{Acknowledgments}
The authors would like to thank Andrea Ferroglia, Thomas Gehrmann and David Kosower for valuable comments on the manuscript, and Ulrich
Haisch, Pietro Falgari and Cedric Studerus for useful comments on the implementation. D.M. wishes
to thank the Swiss national foundation  and the US Department of Energy which supported this work under contracts 200020-109162, PBZH2-117028 and DE-AC02-76SF00515. T.H. acknowledges
hospitality from the Institute for Theoretical Physics of the University of Z\"urich where a part of this work was performed. T.H.\ is
supported by Deutsche Forschungsgemeinschaft, SFB/TR 9 ``Computergest\"{u}tzte Theoretische Teilchenphysik''.

\appendix
\section{Analytic continuation of the expansion}\label{sec:analytic}
The results given by the package are for arguments in the interval (0, 1). When the expansion
has to be evaluated elsewhere in the complex plane, care has to be taken. Let us analyze the three possible types of arguments inside
the HPLs.
\subsection{$H(...,i\sqrt{\frac{x}{1-x}})$}
The expansion of HFs of the types $2^0_1$, $3^0_1$, $2_1^2$, and $3_2^3$ contains terms of the form
\[H\left(...,i\sqrt{\frac{x}{1-x}}\right) \]
as, for example in
\be
{}_2F_1\left(\frac{1}{2}+\epsilon,\frac{1}{2};\frac{3}{2},x\right)=-\frac{i}{2\sqrt{x}}\left(
H_+\left(i\sqrt{\frac{x}{1-x}}\right)
-\epsilon H_{+-}\left(i\sqrt{\frac{x}{1-x}}\right)
+{\cal O}\left(\epsilon^2\right)
\right)\;. 
\ee
The function
\[z\to i\sqrt{\frac{z}{1-z}}\]
in the argument of the HPL maps the complex plane into the upper half plane, with a branch cut starting at $z=1$ and following the real axis to $+\infty$, as does the hypergeometric function. For $x<0$, the argument of the HPL is real, but there is a sign ambiguity due to the convention one uses to define the square root of a negative number. Since HPLs with an odd number of $+$ weights are odd, this ambiguity is removed by the factor of $\sqrt{x}$ accompanying the HPL, provided the same convention for the sign of the square root is used.      
\subsection{$H(...,\sqrt{x})$}
The expansion of HFs of the types $2_1^1$, $3_1^1$, $3_2^2$, $4_1^1$, and $4^3_3$ contains terms of the form
\[H\left(...,\sqrt{x}\right). \]
for $x<0$, the argument of the HPL is complex. HPLs of complex arguments with $\pm$ weights are either purely imaginary or purely
real\footnote{provided there is no 0 weight on the right of the index vector, which is the case in the expansion we deal with.}
depending on the number of + weights in the index vector. The HPLs that are purely complex are accompanied by factors of $\sqrt{x}$ and
the HPLs that are real are not, so that the expansion is also real for $x<0$ and, again, the sign ambiguity is removed if the same
convention for the square root of a negative number is used both in the argument of the HPL and in its prefactor.      
\subsection{$H(...,\frac{1-\sqrt{1-x}}{1+\sqrt{1-x}})$}
The expansion of HFs of the types $2_0^1$ and $3_0^1$ contains terms of the form
\[H\left(...,\frac{1-\sqrt{1-x}}{1+\sqrt{1-x}}\right). \]
In this case, the HPLs are more conveniently expressed with integer weights. The branch cut of the argument of the HPLs for $x>1$ corresponds to that of the hypergeometric function. There the argument of the HPL is on the unit circle and the HPL develops an imaginary part, as does the HF.        
\section{Computation of the basis functions}\label{sec:basis}
Since the integration and differentiation operators introduced above with appropriate coefficients allow one to go from one HF to
another of the same type $P_s^r$, the choice of which HF should be the basis function is arbitrary. In this section, we show the result
of the application of the strategy described in section~\ref{sec:generalstrat} to functions ${}_PF_{P-1}$ of types $2^1_0$, $3^1_1$,
$3^0_1$, and $3^2_2$. We will finally show in section~\ref{sec:basis312} that in the expansion of HFs of type $3^1_2$ new weights have
to be introduced. Some of the expansions in this section could have been also computed with other methods like for example those presented in
Refs.~\cite{Weinzierl:2004bn,Kalmykov:2006pu,Kalmykov:2006hu}, but our method yields the expansion directly in a form suitable to be used in the algorithm described above.
\subsection{HFs of type $2^1_0$}
For  the expansion of the basis function  $B^1_0(\{0,0;1\},\{\alpha_1,\alpha_2;\beta_1\})$ we consider the expansion of the HF
\begin{equation}
{}_2F_1\left(\frac{1}{2}+\alpha_1\epsilon,\alpha_2\epsilon;1+\beta_1\epsilon;x\right)\;,
\end{equation}
for which we found the expansion
\be
{}_2F_1\left(\frac{1}{2}+\alpha_1\epsilon,\alpha_2\epsilon;1+\beta_1\epsilon;x\right)=
1+\sum\limits_{j=1}^\infty \epsilon^j\sum\limits_{l\in \{-1,0,+1\}^j}c(l) H_l\left(\omega\right)\;.
\ee
The coefficient functions $c$ are found recursively by applying the following rules.
\begin{eqnarray}
c(0)&=&c(1)=0\; ,\\
c(-1)&=&2 \alpha_2\; ,\\
c(1,...)&=&0\; ,\\
c(-1,...)&=&2\alpha_2 c(...)\; ,\\
c(0,-1,...)&=&2\alpha_2 (\alpha_1-\alpha_2) c(...) +2  \alpha_1 c(0,...) \; ,\\
c(0,1,...)&=&2(\alpha_1+\alpha_2-\beta_1) \left(\alpha_2 c(...)+c(0,...)\right) \; ,\\
c(0,0,...)&=&-\beta_1  c(0,...)\; ,
\end{eqnarray}
where
\be
\omega=\frac{1-\sqrt{1-x}}{1+\sqrt{1-x}}\, .
\ee
\subsection{HFs of  type $3^1_1$ }
We consider the function 
\[{}_3F_2(\frac{1}{2}+\alpha_1\epsilon,1+\alpha_2\epsilon,1+\alpha_3\epsilon;\frac{1}{2}+\beta_1\epsilon,1+\beta_2\epsilon;x)\;.\]
The expansion of this function is best written as a function of the parameters
\begin{eqnarray}\label{eq:Bparams}
s&=&\alpha_2+\alpha_3-\beta_2\; ,\nonumber\\
d_1&=&\alpha_2-\beta_2\; ,\nonumber\\
d_2&=&\alpha_3-\beta_2\; ,\nonumber\\
d_3&=&\alpha_1-\beta_1.
\end{eqnarray}
Like in the example in section \ref{sec:simpleexample} we make the ansatz
\begin{eqnarray}
\lefteqn{{}_3F_2(\frac{1}{2}+\alpha_1\epsilon,1+\alpha_2\epsilon,1+\alpha_3\epsilon;\frac{1}{2}+\beta_1\epsilon,1+\beta_2\epsilon;x)=}&&\nonumber\\
&&\frac{1}{1-x}\left(1+\sum\limits_{j=1}^\infty \epsilon^j\sum\limits_{l\in \{+,0,-\}^j}c(l;x) H_l(\sqrt{x})\right)\;.
\end{eqnarray}
The coefficient functions are found recursively by applying the following rules
\begin{eqnarray}
c(l;x)&=&c(l)\times\left\{\begin{array}{cc}\sqrt{x},\quad &\textnormal{odd number of + in $l$}\\1&\textnormal{even number of + in $l$}\end{array}\right.\\
c(+)&=&d_3,\quad c(0)=0,\quad c(-)=s\;.
\end{eqnarray}
The prefactor for $+$ weights makes sure that the expansion is real, as adding a ``+" in the weight changes a real HPL into an imaginary one. 
\begin{eqnarray}
\lefteqn{c(+,w_1,...,w_n)=}&&\nonumber\\
&&c(w_1,...,w_n)\left\{\begin{array}{cc} s &\textnormal{odd number of + in $ \{w_1,...,w_n \}$}\\
d_3&\textnormal{even number of + in $ \{w_1,...,w_n \}$ }
\end{array}\right.\nonumber\\
\lefteqn{c(0,w_1,...,w_n)=}&&\nonumber\\
&&c(w_1,...,w_n)\left\{\begin{array}{cc} -2 \beta_2 &\textnormal{odd number of + in $ \{w_1,...,w_n \}$}\\
-2 \beta_1&\textnormal{even number of + in $ \{w_1,...,w_n \}$ and $w_1=0$}\\
\displaystyle\frac{2 d_1 d_2}{s}&\textnormal{even number of + in $ \{w_1,...,w_n \}$ and $w_1\not=0$}
\end{array}\right.\nonumber\\
\lefteqn{c(-,w_1,...,w_n)=}&&\nonumber\\
&&c(w_1,...,w_n)\left\{\begin{array}{cc}d_3 &\textnormal{odd number of + in $ \{w_1,...,w_n \}$}\\
s &\textnormal{even number of + in $ \{w_1,...,w_n \}$ }\;.
\end{array}\right.
\end{eqnarray}
The basis function is then found by subtracting unity from the HF.
\be\
B^1_1(\{0,1,1;0,1\},\{\alpha_1,\alpha_2,\alpha_3;\beta_1,\beta_2\})=\frac{1}{1-x}\left(x+\sum\limits_{j=1}^\infty \epsilon^j\sum\limits_{l\in \{+,0,-\}^j}c(l;x) H_l(\sqrt{x})\right)\;.
\ee
The expansion of the basis function
$B_1^1(\{0,1;0\},\{\alpha_1,\alpha_2;\beta_1\})$ is recovered by setting
 \[s\rightarrow \alpha_2,\quad d_2\rightarrow 0,\quad d_3\to\alpha_1-\beta_1,\]
and that of the basis function $B_0^0(\{1,1;1\},\{\alpha_1,\alpha_2;\beta_1\})$  by taking the limit
\[s\rightarrow \alpha_1+\alpha_2-\beta_1,\quad d_1\rightarrow \alpha_1-\beta_1,\quad d_2\rightarrow \alpha_2-\beta_1,\quad d_3\rightarrow 0\;.\]
We see that in this limit all HPLs with weight + disappear from the expansion. Since only HPLs with weight 0 and $-$ can be transformed
into HPLs of argument $x^2$ (see \cite{Maitre:2007kp}) we expect only such HPLs to be present in the expansion of HFs with only integer
parameters, since we know that the expansion of these HFs can be written in terms of HPLs of argument $x$~\cite{Kalmykov:2007pf}.    
\subsection{HFs of the type $3^0_1$}
We consider the HF ${}_3F_2(\alpha_1\epsilon,1+\alpha_2\epsilon,1+\alpha_3\epsilon;\frac{1}{2}+\beta_1\epsilon,1+\beta_2\epsilon;x)$ and make the ansatz
\begin{eqnarray}
\lefteqn{{}_3F_2(\alpha_1\epsilon,1+\alpha_2\epsilon,1+\alpha_3\epsilon;\frac{1}{2}+\beta_1\epsilon,1+\beta_2\epsilon;x)=}&&\nonumber\\
&&1+\sum\limits_{j=1}^\infty \epsilon^j\sum\limits_{l\in \{+,0,-\}^j}c(l;x) H_l\left(i\sqrt{\frac{x}{1-x}}\right)\;.
\end{eqnarray}
The coefficient functions are found recursively by applying the following rules
\begin{eqnarray}
\lefteqn{c(l;x)=c(l)\times\left\{\begin{array}{cc}i\sqrt{\frac{x}{1-x}},\quad &\textnormal{odd number of + in $l$}\\1&\textnormal{even number of + in $l$}\end{array}\right.}\\
\lefteqn{c(+)=-\alpha_3,\quad c(0)=0,\quad c(-)=0 \, ,}&& \\
\lefteqn{c(+,w_1,...,w_n)=c(w_1,...,w_n)}&&\nonumber\\
&\times&\left\{\begin{array}{cc} s &\textnormal{odd number of + in $ \{w_1,...,w_n \}$}\\
\displaystyle\frac{d_1 d_2-s(d_3+\beta_1)}{s}&\textnormal{even number of + in $ \{w_1,...,w_n \}$ and $w_1=+$}\\
s-d_1-d_2-d_3+\beta_1&\textnormal{even number of + in $ \{w_1,...,w_n \}$ and $w_1\not=+$}
\end{array}\right.\nonumber\\
\lefteqn{c(0,w_1,...,w_n)=c(w_1,...,w_n)}&&\nonumber\\
&\times&\left\{\begin{array}{cc} -2 \beta_2 &\textnormal{odd number of + in $ \{w_1,...,w_n \}$}\\
\displaystyle\frac{2 d_1 d_2}{s}&\textnormal{even number of + in $ \{w_1,...,w_n \}$ and $w_1=+$}\\
2 (d_1+d_2-s)&\textnormal{even number of + in $ \{w_1,...,w_n \}$ and $w_1\not=+$}
\end{array}\right.\nonumber\\
\lefteqn{c(-,w_1,...,w_n)=c(w_1,...,w_n)}&&\nonumber\\
&\times&\left\{\begin{array}{cc} -s-\alpha_3 &\textnormal{odd number of + in $ \{w_1,...,w_n \}$}\\
\displaystyle\frac{d_1 d_2}{s}&\textnormal{even number of + in $ \{w_1,...,w_n \}$ and $w_1=+$}\\
d_1+d_2-s&\textnormal{even number of + in $ \{w_1,...,w_n \}$ and $w_1\not=+$}\;,
\end{array}\right.\nonumber\\
\end{eqnarray}
where we used the definitions of (\ref{eq:Bparams}). The basis function \[B_1^0(\{0,1,1;0,1\},\{\alpha_1,\alpha_2,\alpha_3;\beta_1,\beta_2\})\]
 is then simply 
\be\label{eq:3B01}
B_1^0(\{0,1,1;0,1\},\{\alpha_1,\alpha_2,\alpha_3;\beta_1,\beta_2\})=\sum\limits_{j=1}^\infty \epsilon^j\sum\limits_{l\in
\{+,0,-\}^j}c(l;x) H_l\left(\sqrt{\frac{x}{x-1}}\right)\;.
\ee
The expression for $B_1^0(\{0,1;0\},\{\alpha_1,\alpha_2;\beta_1\})$ is found by setting
$\alpha_3 \rightarrow \beta_2$
in (\ref{eq:3B01}).
\subsection{HFs of type $3^2_2$}
We consider the HF ${}_3F_2(\frac{1}{2}+\alpha_1\epsilon,\frac{1}{2}+\alpha_2\epsilon,1+\alpha_3\epsilon;\frac{1}{2}+\beta_1\epsilon,\frac{1}{2}+\beta_2\epsilon;x)$ and make the ansatz
\begin{eqnarray}
\lefteqn{{}_3F_2(\frac{1}{2}+\alpha_1\epsilon,\frac{1}{2}+\alpha_2\epsilon,1+\alpha_3\epsilon;\frac{1}{2}+\beta_1\epsilon,\frac{1}{2}+\beta_2\epsilon;x)=}&&\nonumber\\
&&\frac{1}{1-x}\left(1+\sum\limits_{j=1}^\infty \epsilon^j\sum\limits_{l\in \{+,0,-\}^j}c(l;x) H_l\left(i\sqrt{\frac{x}{1-x}}\right)\right)\;.
\end{eqnarray}
The expansion of this function is best written as a function of the parameters
\begin{eqnarray}\label{eq:B22params}
s&=&\alpha_1+\alpha_2-\beta_1-\beta_2\; ,\nonumber\\
s_\beta&=&\beta_1+\beta_2\; ,\nonumber\\
p_\alpha&=&\alpha_1\alpha_2\; ,\nonumber\\
p_\beta&=&\beta_1\beta_2\;.
\end{eqnarray}
The coefficient functions are found recursively by applying the following rules
\begin{eqnarray}
c(l;x)&\equiv&c(l)\times\left\{\begin{array}{cc}i\sqrt{\frac{x}{1-x}},\quad &\textnormal{odd number of + in $l$}\\1&\textnormal{even number of + in $l$}\end{array}\right.\nonumber \\
c(+)&=&s,\quad c(0)=0,\quad c(-)=0  
\end{eqnarray}
\begin{eqnarray}
\lefteqn{c(+,w_1,...,w_n)=}&&\nonumber\\
&&c(w_1,...,w_n)\times\left\{\begin{array}{cc} 
\alpha_3 &\textnormal{odd number of + in $ \{w_1,...,w_n \}$}\\
s        &\textnormal{even number of + in $ \{w_1,...,w_n \}$}\\
\end{array}\right.\nonumber\\
\lefteqn{c(0,w_1,...,w_n)=}&&\nonumber\\
&&\left\{\begin{array}{cl}
-2s_\beta c(w_1,...,w_n)-4p_\beta c(w_2,...,w_n) & \textnormal{odd number of + in $ \{w_1,...,w_n \}$}\\
0&\textnormal{\rule{-0.75cm}{0cm}even number of + in $ \{w_1,...,w_n \}$ }
\end{array}\right.\nonumber\\
\lefteqn{c(-,w_1,...,w_n)=c(w_1,...,w_n)}&&\nonumber\\
&&\left\{\begin{array}{cc}
s &      \textnormal{odd number of + in $ \{w_1,...,w_n \}$}\\
\alpha_3&\textnormal{even number of + in $ \{w_1,...,w_n \}$\;. }
\end{array}\right.
\end{eqnarray}
The basis function $B_2^2(\{0,0,1;0,0\},\{\alpha_1,\alpha_2,\alpha_3;\beta_1,\beta_2)\}$ is then simply 
\begin{eqnarray}\label{eq:3B22}
\lefteqn{B_2^2(\{0,0,1;0,0\},\{\alpha_1,\alpha_2,\alpha_3;\beta_1,\beta_2\})=}\nonumber\\
&&\frac{1}{1-x}\left(x+\sum\limits_{j=1}^\infty \epsilon^j\sum\limits_{l\in \{+,0,-\}^j}c(l;x) H_l\left(i\sqrt{\frac{x}{1-x}}\right)\right)\;.
\end{eqnarray}
The results of Eq.~(\ref{eq:ruleB11}) are recovered by setting either $\alpha_1$ or $\alpha_2$ equal to one of the $\beta$'s. 
\subsection{HFs of type $3^1_2$}\label{sec:basis312}
For  the expansion of the basis function  $B^1_2(\{0,0,1;0,0\},\{\alpha_1,\alpha_2,\alpha_3;\beta_1,\beta_2\})$ we consider the expansion of the HF
\begin{equation}
{}_3F_2(\frac{1}{2}+\alpha_1\epsilon,\alpha_2\epsilon,1+\alpha_3\epsilon;\frac{1}{2}+\beta_1\epsilon,\frac{1}{2}+\beta_2\epsilon;x)\;,
\end{equation}
for which we found the expansion
\begin{eqnarray}
\lefteqn{{}_3F_2\left(\frac{1}{2}+\alpha_1\epsilon,\alpha_2\epsilon,1+\alpha_3\epsilon;\frac{1}{2}+\beta_1\epsilon,\frac{1}{2}+\beta_2\epsilon;x\right)=}&&\nonumber\\
&&1-\alpha_2 \epsilon \omega  H(+;\omega ) \nonumber\\
&-&\alpha_2 \epsilon^2 \big((-\left(\alpha_2+\alpha_3\right)\omega  H(-,+;\omega )  +\alpha_3H(+,+;\omega )\nonumber\\
&&\qquad\qquad +2\left(\alpha_1-\beta _1-\beta _2\right) \omega  H(0,+;\omega ) \big)\nonumber\\
&+&\alpha_2\epsilon^3\bigg( 
\alpha_2\alpha_3 \omega  H(+,+,+;\omega )
- \left(\alpha_2+\alpha_3\right)^2\omega  H(-,-,+;\omega )\nonumber\\ 
&&\qquad-4  \left(\alpha_1-\beta _1-\beta _2\right)^2\omega  H(0,0,+;\omega ) 
+\alpha_3\left(\alpha_2+\alpha_3\right) H(+,-,+;\omega ) \nonumber\\
&&\qquad+2 \left(\alpha_2+\alpha_3\right) \left(\alpha_1-\beta _1-\beta _2\right)\omega  H(0,-,+;\omega )
\nonumber\\ 
&&\qquad-2 \alpha_3 \left(\alpha_1-\beta _1-\beta _2\right) H(+,0,+;\omega ) \nonumber\\ 
&& \qquad +2 \left(\alpha_2+\alpha_3\right) \left(\alpha_1-\beta _1-\beta _2\right)\omega  H(-,0,+;\omega)\bigg)\nonumber\\
&+&4 \epsilon^3 \alpha_2   \left(\alpha_1-\beta _1\right) \left(\alpha_1-\beta _2\right)\omega  H\left(\frac{1}{2},\frac{1}{2},+;\omega \right) +{\cal O}(\epsilon^4)\;,
\end{eqnarray}
where
\[\omega=i\sqrt{\frac{x}{1-x}}\,.\]
In the last term, we have to define new weights for the HPLs:
\begin{equation}
f_\frac{1}{2}(t)=\frac{1}{t\sqrt{1-t^2}}\;.
\end{equation}
The appearance of this new weight shows that the HPLs of weights $+,-,0$ are not sufficient for the construction of the expansion of all
HFs with half-integer parameters. This is not a limitation to the algorithm described in this paper. However it is a limitation for the current  implemetation, as these weights are not supported by the package HPL.

\end{document}